\documentclass{IEEEtran}
\IEEEoverridecommandlockouts
\usepackage[space]{cite}
\usepackage{authblk}
\usepackage{amsmath, amsthm, amssymb, amsfonts, bm}
\usepackage{mathtools}
\usepackage{tikz}
\usepackage{caption}
\usepackage[skip = 0pt]{subcaption}
\newtheorem{theorem}{Theorem}
\newtheorem{lemma}{Lemma}

\newtheorem{corollary}[theorem]{Corollary}

\newtheorem{proposition}[theorem]{Proposition}
\newtheorem{deff}{Definition}
\usepackage{algorithm}
\usepackage{algorithmic}

\usepackage{array}
\usepackage{graphicx}
\usepackage{textcomp}
\usepackage{xcolor}
\usepackage{adjustbox}
\usepackage{fancyhdr}

\usepackage{xr-hyper}
\definecolor{peachfill}{RGB}{255, 230, 204}
\definecolor{darkorangeborder}{RGB}{215, 155, 0}
\definecolor{pinkfill}{RGB}{248, 206, 204}
\definecolor{darkredborder}{RGB}{184, 84, 80}
\definecolor{bluefill}{RGB}{218, 232, 252}
\definecolor{blueborder}{RGB}{108, 142, 191}
\definecolor{greenfill}{RGB}{213, 232, 212}
\definecolor{greenborder}{RGB}{130, 179, 102}
\definecolor{purplefill}{RGB}{225, 213, 231}
\definecolor{purpleborder}{RGB}{150, 115, 166}
\definecolor{yellowfill}{RGB}{255, 242, 204}
\definecolor{yellowborder}{RGB}{214, 182, 86}
\definecolor{palegray}{RGB}{242, 242, 242}

\newcommand\middlescript[1]{\vcenter{\hbox{$\scriptstyle #1$}}}
\newcommand{\norm}[1]{\lVert#1\rVert}
\usepackage{nomencl}
\makenomenclature

\usepackage{hyperref}
\hypersetup{
    colorlinks=true,
    linkcolor=blue,
    filecolor=magenta,      
    urlcolor=cyan
    }

\begin{document}
\bstctlcite{IEEEexample:BSTcontrol}
\setlength\tabcolsep{2.4pt}
\setlength{\headheight}{15.17393pt}

\title{Deep Regularized Compound Gaussian Network for Solving Linear Inverse Problems
}
\author[1, 2]{Carter Lyons\thanks{carter.lyons@colostate.edu}
}
\author[1]{Raghu G. Raj\thanks{raghu.g.raj@ieee.org}
}
\author[2]{Margaret Cheney\thanks{margaret.cheney@colostate.edu}\thanks{This work was sponsored in part by the Office of Naval Research via the NRL base program and under award number N00014-21-1-2145 and in part by the Air Force Office of Scientific Research under award number FA9550-21-1-0169.}\thanks{Project codes are accessible at \url{https://github.com/clyons19/DR-CG-Net.}}
}
\affil[1]{U.S. Naval Research Laboratory, Washington, D.C.}
\affil[2]{Colorado State University, Fort Collins, CO}

\maketitle
\thispagestyle{fancy}

\begin{abstract}
Incorporating prior information into inverse problems, e.g. via maximum-a-posteriori estimation, is an important technique for facilitating robust inverse problem solutions. In this paper, we devise two novel approaches for linear inverse problems that permit problem-specific statistical prior selections within the compound Gaussian (CG) class of distributions. The CG class subsumes many commonly used priors in signal and image reconstruction methods including those of sparsity-based approaches. The first method developed is an iterative algorithm, called generalized compound Gaussian least squares (G-CG-LS), that minimizes a regularized least squares objective function where the regularization enforces a CG prior. G-CG-LS is then unrolled, or unfolded, to furnish our second method, which is a novel deep regularized (DR) neural network, called DR-CG-Net, that learns the prior information. A detailed computational theory on convergence properties of G-CG-LS and thorough numerical experiments for DR-CG-Net are provided. Due to the comprehensive nature of the CG prior, these experiments show that DR-CG-Net outperforms competitive prior art methods in tomographic imaging and compressive sensing, especially in challenging low-training scenarios.
\end{abstract}

\begin{IEEEkeywords}
Machine learning, deep neural networks, inverse problems, nonlinear programming, least squares methods
\end{IEEEkeywords}
\vspace*{-0.5cm}
\nomenclature[01]{$\mathbb{R}$}{Set of real numbers.}
\nomenclature[02]{$\bm{v}$}{$=[v_i]\in\mathbb{R}^n$. Boldface characters are vectors.}
\nomenclature[03]{$A$}{$ = [A_{ij}]\in\mathbb{R}^{m\times n}$. Uppercase characters are matrices.}
\nomenclature[04]{$(\cdot)^T$}{Transpose of vector or matrix $(\cdot)$.}
\nomenclature[05]{$\odot$}{Hadamard product.}
\nomenclature[06]{$A_{\bm{v}}$}{$=A\textnormal{Diag}(\bm{v})$ for $A$ and $\bm{v}$ of compatible size.}
\nomenclature[07]{$\mathcal{P}_{\mathcal{C}}$}{Unique projection operator onto convex set $\mathcal{C}$.}
\printnomenclature

\section{Introduction}

\IEEEPARstart{L}{inear} inverse problems enjoy extensive applications throughout scientific and engineering disciplines -- including X-ray computed tomography, magnetic resonance imaging (MRI), radar imaging, sonar imaging, and compressive sensing (CS) -- making it an active area of research.

Until recently, iterative or model-based approaches were the main methodologies to solve (linear) inverse problems and typically incorporate an expected prior density into an estimation algorithm. Example iterative algorithms include the Iterative Shrinkage and Thresholding Algorithm (ISTA)~\cite{ISTA, fast_ISTA}, Bayesian Compressive Sensing~\cite{BCS}, and Compressive Sampling Matching Pursuit~\cite{Compressive_Sensing, CoSamp}. Many previous works implement a generalized Gaussian prior~\cite{fast_ISTA, ISTA, CoSamp}, which is encompassed by the broad compound Gaussian (CG) prior that better captures statistical properties of images~\cite{Scale_Mixtures, Wavelet_Trees, CGNetTSP, chance2011information, waveform_opt}.

Innovations in artificial intelligence (AI) and machine learning (ML), particularly deep learning, paired with advancements in computational efficiency, have yielded prominent tools with tremendous applications to signal and image processing problems including inverse problems. Deep learning approaches construct a function mapping between a domain and range of interest by training a deep neural network (DNN), i.e. a NN with a sizeable number of layers, with instantiations of domain-range pairs. For inverse problems, measurement-signal pairs serve as the domain-range pairs used in training. Many previous works use standard DNN methods including convolutional neural networks (CNN), recurrent neural networks, or generative adversarial networks~\cite{reconnet, jin2017FBPConvNet, he2020iRadonMAP, ganbora, liang2020deep, wang2020deep}. 

Algorithm unrolling or unfolding, stemming from the original work of Gregor and LeCun~\cite{learned_ISTA}, is a recent approach to inverse problems that combines the model-based and data-driven methods by structuring the layers of a DNN upon an iterative algorithm.  While a standard DNN acts as a black-box process, we have some understanding of the inner workings of an unrolled DNN by understanding the original iterative algorithm. Additionally, algorithm unrolling seamlessly allows for the incorporation of a prior density into a deep learning framework. These factors contribute to the success of algorithm unrolled methods, which have shown excellent performance in image estimation while offering interpretability of the network layers~\cite{algorithm_unrolling}. Algorithm unrolling has been applied to many iterative algorithms including ISTA~\cite{algorithm_unrolling, MADUN, zhang2018ista, learned_ISTA, xiang2021FISTANet}, proximal gradient or gradient descent~\cite{learned_proximal_operators, zhang2022learn++}, the inertial proximal algorithm for non-convex optimization~\cite{su2020iPianoNet}, and the primal-dual algorithm~\cite{adler2018LPD}. 

\subsection{Contributions}
This paper constructs a CG-based iterative signal estimation algorithm and then leverages algorithm unrolling to develop novel DNN structures, for solving linear inverse problems, that are fundamentally informed by a learned CG prior. In particular we:
\begin{enumerate}
    \item Develop a novel CG-based iterative algorithm, called generalized compound Gaussian least squares (G-CG-LS), for linear inverse problems, which reduces to previous CG-based approaches~\cite{CGNetTSP} and Tikhonov regularization as special cases. Specifically, G-CG-LS solves a regularized least squares optimization problem with two regularization terms consisting of a Gaussian term and an implicitly defined non-Gaussian term, which together enforce a CG prior.
    \item Derive a theoretical convergence analysis for G-CG-LS to stationary points of the underlying cost function to provide insight into the performance of G-CG-LS.
    \item Construct novel CG-based DNNs, called deep regularized compound Gaussian network (DR-CG-Net), by applying the algorithm unrolling technique to G-CG-LS, such that the implicit portion of the regularization is trainable, thereby allowing for learning of the prior density within the powerful CG class of distributions.
    \item Present novel empirical results for DR-CG-Net on tomographic imaging and CS problems. We demonstrate the effectiveness of DR-CG-Net when limited to small training datasets and show a significant improvement over other state-of-the-art deep learning-based approaches.
\end{enumerate}
Our proposed G-CG-LS and DR-CG-Net are expansions on prior art CG-based methods~\cite{APSIPAlyonsrajcheney, Asilomarlyonsrajcheney, CGNetTSP}. A core novelty of the DR-CG-Net framework, compared to previous CG-based methods, is the ability to learn arbitrary signal priors within the CG distribution class. As such, we generalize to a broader signal prior within the CG-prior context. Furthermore, our proposed methods can be reduced to prior CG-based methods~\cite{APSIPAlyonsrajcheney, Asilomarlyonsrajcheney, CGNetTSP} as special cases. Empirically, we demonstrate significant improvement, in estimated signal quality and time, for DR-CG-Net over previous CG-based methods.

\section{Background}

In this section, we provide brief introductions to the CG prior, DNNs, and algorithm unrolling as these are the key components we combine to furnish the methods developed in this paper. First, the linear measurement (i.e. forward) model we consider throughout this paper is
\begin{equation}
    \bm{y} = A\bm{c} + \bm{\nu} \label{eqn:linear_msrmt}
\end{equation}
where $\bm{y}\in\mathbb{R}^m$ are the measurements, $A\in\mathbb{R}^{m\times n}$ is a sensing matrix, $\bm{c}\in\mathbb{R}^n$ is the unknown signal, and $\bm{\nu}\in\mathbb{R}^m$ is additive white noise. In many applications of interest, $m\ll n$ and $A$ is decomposed as $A = \Psi\Phi$ for $\Psi\in\mathbb{R}^{m\times n}$ a measurement matrix and $\Phi\in\mathbb{R}^{n\times n}$ a change of basis dictionary. This formulation results from representing an original signal, $\bm{s}$, with respect to (w.r.t.) $\Phi$ as $\bm{s} = \Phi\bm{c}$. Signal reconstruction, or estimation, is the inverse problem of recovering $\bm{c}$, or $\bm{s}$, given $\bm{y}$, $\Psi$, and $\Phi$.

 \subsection{Compound Gaussian Prior}
 
  A fruitful way to formulate inverse problems is by Bayesian estimation, in particular, a maximum a posteriori (MAP) estimate. The MAP estimate of $\bm{c}$ from (\ref{eqn:linear_msrmt}) depends crucially on the assumed prior density of $\bm{c}$, which serves to incorporate domain-level knowledge into the inverse problem. Many previous works employ a generalized Gaussian prior~\cite{Scale_Mixtures, Wavelet_Trees} such as a Gaussian prior for Tikhonov regression~\cite{bertero1988linear} or a Laplacian prior as is predominant in the CS framework~\cite{Compressive_Sensing, ISTA, fast_ISTA, CoSamp}. 

Through the study of the statistics of image sparsity coefficients, it has been shown that coefficients of natural images exhibit self-similarity, heavy-tailed marginal distributions, and self-reinforcement among local coefficients~\cite{Wavelet_Trees}. Such properties are not encompassed by the generalized Gaussian prior. Instead, CG densities~\cite{HB-MAP}, also known as Gaussian scale mixtures~\cite{Scale_Mixtures, Wavelet_Trees}, better capture these statistical properties of natural images and images from other modalities such as radar~\cite{chance2011information, waveform_opt}. 
A useful formulation of the CG prior lies in modeling a signal $\bm{c}$ as the Hadamard product
\begin{align}
    \bm{c} &= \bm{z}\odot\bm{u} \coloneqq [z_1u_1, z_2u_2, \ldots, z_nu_n]^T \label{eqn:CG}
\end{align}
such that $\bm{u}\sim \mathcal{N}(\bm{0},\Sigma_u)$, $\bm{z}\sim p_{\bm{z}}$ is a positive random vector, and $\bm{u}$ and $\bm{z}$ are independent~\cite{Wavelet_Trees,HB-MAP}. We call $\bm{u}$ and $\bm{z}$ as the Gaussian variable and scale variable, respectively. It has been shown in~\cite{Wavelet_Trees, CGNetTSP} that by suitably defining the distribution of $\bm{z}$, the CG prior subsumes many well-known distributions including the generalized Gaussian, student's $t$, $\alpha$-stable, and symmetrized Gamma distributions.

Previously, the CG prior has been used with a log-normal distributed scale variable in an iterative MAP estimate of wavelet and discrete cosine transformation (DCT) image coefficients~\cite{HB-MAP, fast_HB-MAP}. Additionally, a log-normal distribution for $\bm{z}$ was used in an alternative and simplified iterative MAP estimation of image coefficients that was subsequently unrolled into a DNN~\cite{CGNetTSP, APSIPAlyonsrajcheney, Asilomarlyonsrajcheney}.  Furthermore, the CG prior using a real-valued random scale variable has been successfully used for image denoising~\cite{portilla2003image} and hyperspectral image CS~\cite{huang2021deep}.

\subsection{Deep Neural Networks and Algorithm Unrolling}

A DNN is a collection of ordered layers, denoted $\bm{L}_0, \bm{L}_1, \ldots, \bm{L}_K$ for $K > 1$, that form a directed acyclic graph starting with the input layer, $\bm{L}_0$, and ending at the output layer, $\bm{L}_K$. Intermediate layers $\bm{L}_1, \ldots, \bm{L}_{K-1}$ are known as hidden layers. Each layer, $\bm{L}_k$, contains $d_k$ nodes, or hidden units~\cite{goodfellow2016deep}, and for simplicity, we identify the nodes with the computed values assigned when a signal is processed by the DNN.

A function $\bm{f}_k:\mathbb{R}^{d_{i_1(k)}}\times \cdots \times \mathbb{R}^{d_{i_j(k)}} \to \mathbb{R}^{d_k}$, that is parameterized by some $\bm{\theta}_k$, defines the computation, i.e. signal transmission, at layer $\bm{L}_k$ where $\mathcal{I}_k\coloneqq\{i_1(k),\ldots, i_j(k)\}\subseteq \{0,1,\ldots, K-1\}$ are the indices of layers that feed into $\bm{L}_k$. That is, given an input signal $\overline{\bm{y}}\in\mathbb{R}^{d_0}$ assigned to $\bm{L}_0$, a DNN is a composition of parameterized functions where
 \[
\bm{L}_k \equiv \bm{f}_k\left(\bm{L}_{i_1(k)}, \ldots, \bm{L}_{i_j(k)}; \bm{\theta}_k\right) \in\mathbb{R}^{d_k}.
\]

Standard convolutional neural networks, as an example, use $\mathcal{I}_k = \{k-1\}$ and $\bm{f}_k(\bm{L}_{k-1}; \bm{\theta}_k = [W_k,\bm{b}_k]) = \sigma(W_k\star \bm{L}_{k-1}+\bm{b}_k)$ where $\bm{b}_k$ is an added bias, $\sigma$ is an activation function applied componentwise, and $W_k\star \bm{L}_{k-1}$ denotes the convolution of the weight kernel $W_k$ and input signal $\bm{L}_{k-1}$.

A DNN learns its parameters, $\bm{\Theta} = (\bm{\theta}_1, \ldots, \bm{\theta}_K)$, by optimizing a loss function, $\mathcal{L}(\bm{\Theta})$, over a training dataset $\mathcal{D} = \{(\overline{\bm{y}}_i, \overline{\bm{c}}_i): i = 1, 2, \ldots, N_s\}$ where each $(\overline{\bm{y}}_i,\overline{\bm{c}}_i)$ satisfies equation (\ref{eqn:linear_msrmt}).  Let $\widehat{\bm{c}}(\overline{\bm{y}}_i;\bm{\Theta})$ denote the DNN output given the input $\overline{\bm{y}}_i$ and parameterization $\bm{\Theta}$. Then the loss function is often defined as the average $\mathcal{L}(\bm{\Theta}) \coloneqq \frac{1}{N_s}\sum_{i = 1}^{N_s} L\left(\widehat{\bm{c}}(\overline{\bm{y}}_i;\bm{\Theta}), \overline{\bm{c}}_i\right)$ where $L\left(\widehat{\bm{c}}(\overline{\bm{y}}_i;\bm{\Theta}), \overline{\bm{c}}_i\right)$ is the loss, or error, between the network output, $\widehat{\bm{c}}(\overline{\bm{y}}_i;\bm{\Theta})$, and the actual coefficients, $\overline{\bm{c}}_i$. Common loss functions for image reconstruction neural networks include mean-squared error (MSE), mean-absolute error (MAE), peak signal-to-noise ratio (PSNR), or structural similarity index measure (SSIM)~\cite{NN_loss}. After training the DNN, the cascade of function compositions collectively models a desired transformation from an input space to an output space.

Algorithm unrolling creates a DNN by assigning the operations from each step, $k$, of any iterative algorithm as the function, $\bm{f}_k$, defining layer $k$. That is, layer $k$ in the DNN corresponds to the output of $k$ iterations of the original iterative algorithm. Then parameters, $\bm{\theta}_k$, on each step, $k$, of the iterative algorithm parameterize $\bm{f}_k$ in the DNN~\cite{learned_ISTA, algorithm_unrolling}. In training the unrolled DNN, each $\bm{\theta}_k$ is learned, which optimizes the iterative algorithm.

\section{Iterative Algorithm: G-CG-LS} \label{sec:interative_alg}

In order to solve the linear inverse problem to (\ref{eqn:linear_msrmt}) for the unknown vector $\bm{c}$, we incorporate the CG prior into $\bm{c}$ via (\ref{eqn:CG}) and perform the following optimization w.r.t. both the Gaussian, i.e. $\bm{u}$, and non-Gaussian, i.e. $\bm{z}$, components
  \begin{align}
          \begin{bmatrix}
  \bm{u}^* &  \bm{z}^*
  \end{bmatrix} &= \underset{(\bm{u},\,\bm{z})\in\mathbb{R}^n\times \mathfrak{Z}}{\arg\min}\,\, F(\bm{u},\bm{z})  \label{eqn:iterative_cost_func}
    \end{align}
    where
    \begin{align}
    F(\bm{u},\bm{z}) \coloneqq \frac{1}{2}\norm{\bm{y}-A(&\bm{z}\odot\bm{u})}_2^2 + \frac{1}{2}\bm{u}^T P_u^{-1}\bm{u} + \mathcal{R}(\bm{z}) \label{eqn:cost function}
\end{align}
such that $\frac{1}{2}\norm{\bm{y}-A(\bm{z}\odot\bm{u})}_2^2$ is defined as the data fidelity term, $P_u\propto \Sigma_u$, $\mathcal{R}(\bm{z})\propto \log(p_{\bm{z}}(\bm{z}))$, and $\mathfrak{Z}\subseteq [0,\infty)^n$ is the domain of $\mathcal{R}$, which we assume to be convex. Our solution for the linear inverse problem to (\ref{eqn:linear_msrmt}) is then given as $\bm{c}^* = \bm{u}^*\odot\bm{z}^*$.

Note that for additive white noise in (\ref{eqn:linear_msrmt}), the estimate in (\ref{eqn:iterative_cost_func}) is equivalent to a MAP estimate (see Proposition~\ref{prop:MAP_estimate} in the Appendix). Our G-CG-LS algorithm, given in Algorithm~\ref{alg:CG-LS}, is an iterative method to approximately solve (\ref{eqn:iterative_cost_func}) through block coordinate descent~\cite{wright2015coordinate}.

To detail Algorithm~\ref{alg:CG-LS}, define $A_{\bm{z}} = A\textnormal{Diag}(\bm{z})$ and note that the minimization of (\ref{eqn:cost function}) w.r.t.~$\bm{u}$ is a Tikhonov regularization, or ridge regression, problem with minimizer $\mathcal{T}(\bm{z}) \equiv \mathcal{T}(\bm{z}; P_u)$ defined as
\begin{align}
   \mathcal{T}(\bm{z}) \coloneqq \underset{\bm{u}}{\arg\min}\,\, F(\bm{u},\bm{z}) &= (A_{\bm{z}}^TA_{\bm{z}} + P_u^{-1})^{-1}A_{\bm{z}}^T\bm{y}. \label{eqn:Tikhonov Solution}
\end{align}
which we call the Tikhonov solution.
For minimizing (\ref{eqn:cost function}) w.r.t. $\bm{z}$, a number, $J$, of descent steps are iteratively applied. Let $g(\bm{z}, \bm{u}):\mathbb{R}^n\times \mathbb{R}^n\to\mathbb{R}^n$ be a descent function of (\ref{eqn:cost function}) w.r.t. $\bm{z}$; we consider two possibilities for $g$ as discussed in Section~\ref{sec:scale variable update methods}. Then, the G-CG-LS estimate of $\bm{z}$ on descent step $j$ of iteration $k$ and the G-CG-LS estimate of $\bm{u}$ on iteration $k$ are given respectively by
\begin{align*}
    \bm{z}_k^{(j)} = g(\bm{z}_k^{(j-1)}, \bm{u}_{k-1}) \hspace{.25cm} \textnormal{ and } \hspace{.25cm} \bm{u}_k 
    = \mathcal{T}(\bm{z}_k; P_u). 
\end{align*} 
Note that we define $\bm{z}_k^{(J)} = \bm{z}_{k+1}^{(0)} \equiv \bm{z}_k$. Lastly, Algorithm~\ref{alg:CG-LS} is initialized as $\bm{z}_0 = \mathcal{P}_{[0,b]^n}(A^T\bm{y})$ and $\bm{u}_0 = \mathcal{T}(\bm{z}_0)$. Using the rectified linear unit (ReLU) activation function, we remark that $\mathcal{P}_{[0,b]^n}(\bm{x}) = [\textnormal{ReLU}(x_i)-\textnormal{ReLU}(x_i-b)]_{i = 1}^n$ and is applied to eliminate negative values in $\bm{z}_0$ and limit the maximum value in $\bm{z}_0$ for numerical stability in calculating $\bm{u}_0$.

Note that the computational time to calculate (\ref{eqn:Tikhonov Solution}) can be reduced by using the Woodbury matrix identity to express $\mathcal{T}$ as (see Section~\ref{supmat-apndx:tikhonov solution} in the supplementary material)
\begin{align}
    \mathcal{T}(\bm{z}; P_u) = P_u A_{\bm{z}}^T(I+A_{\bm{z}}P_uA_{\bm{z}}^T)^{-1}\bm{y}, \label{eqn:Tikhonov solution rewritten}
\end{align}
which turns the $n\times n$ linear system of equations in (\ref{eqn:Tikhonov Solution}) into an $m\times m$ linear system of equations. Additionally, we remark that (\ref{eqn:cost function}) is strongly convex w.r.t. $\bm{u}$ and thus the Tikhonov solution is efficiently approximated by Nesterov accelerated gradient descent (NAGD)~\cite{nesterov1983method}. Hence, for sufficiently high-dimensional signals of interest, such that the linear solve in (\ref{eqn:Tikhonov solution rewritten}) is infeasible, we replace the exact calculation of (\ref{eqn:Tikhonov solution rewritten}) by a NAGD approximation. To detail the NAGD approximation, define for $\eta > 0$
\begin{align}
    r_u(\bm{u}, \bm{z}) &\coloneqq \bm{u} - \eta\nabla_u F(\bm{u},\bm{z}) \nonumber \\
    &= \bm{u} - \eta\left(A_{\bm{z}}^T(A_{\bm{z}}\bm{u} - \bm{y}) + P_u^{-1}\bm{u}\right). \label{eqn:G-CG-LS GD u step}
\end{align}
The estimate of $\bm{u}$ on descent step $j+1$ of iteration $k$ is then given by
\begin{align*}
    \bm{u}_k^{(j+1)} = r_u(\bm{u}_k^{(j)}, \bm{z}_k) + \beta (r_u(\bm{u}_k^{(j)}, \bm{z}_k) - r_u(\bm{u}_k^{(j-1)}, \bm{z}_k))
\end{align*}
where $\beta\geq 0$ is a momentum control parameter and $\bm{u}_k^{(0)} = \bm{u}_k^{(-1)} = \bm{u}_{k-1}$. Let $J_u$ denote the number of NAGD steps, then the estimate of $\bm{u}$ on iteration $k$ is given as $\bm{u}_k = u_k^{(J_u)}.$ Note that we denote this composition of all $J_u$ NAGD steps as $\widetilde{\mathcal{T}}(\bm{u}, \bm{z})\equiv \widetilde{\mathcal{T}}(\bm{u}, \bm{z}; P_u)$ to emphasize that the NAGD estimate approximates the Tikhonov solution, $\mathcal{T}$. That is, $\bm{u}_k = \bm{u}_{k-1}^{(J_u)} = \widetilde{\mathcal{T}}(\bm{u}_{k-1}, \bm{z}_{k})$.

\begin{algorithm}[!t]
\caption{G-CG-LS}\label{alg:CG-LS}
\begin{algorithmic}[1]
\STATE $\bm{z}_0 = \mathcal{P}_{[0,b]^n}(A^T\bm{y})$ and $\bm{u}_0 = \mathcal{T}(\bm{z}_0)$ (or $\widetilde{\mathcal{T}}(\bm{0}, \bm{z}_0)$ for NAGD) \label{line:initial}
\FOR{$k\in \{1,2,\ldots, K\}$}
\STATE  {\underline{$\bm{z}$ \textsc{estimation}}:}
\STATE  $\bm{z}_k^{(0)} = \bm{z}_{k-1}$
\FOR{$j\in \{1,2, \ldots, J\}$}
\STATE  $\bm{z}_{k}^{(j)} = g(\bm{z}_{k}^{(j-1)}, \bm{u}_{k-1})$ \label{line:z update}
\ENDFOR
\STATE $\bm{z}_{k} = \bm{z}_k^{(J)}$ 
\STATE {\underline{$\bm{u}$ \textsc{estimation}}:}
\STATE $ \bm{u}_{k} = \mathcal{T}(\bm{z}_k)$ (or $\widetilde{\mathcal{T}}(\bm{u}_{k-1}, \bm{z}_k)$ for NAGD)\label{line:u update}
\ENDFOR
\ENSURE{$\bm{c}^* = \bm{z}_K\odot \bm{u}_K$}
\end{algorithmic}
\end{algorithm}

As our G-CG-LS method generalizes previous CG-based iterative algorithms~\cite{APSIPAlyonsrajcheney, Asilomarlyonsrajcheney, CGNetTSP}, we briefly explain key differences. Previous works~\cite{APSIPAlyonsrajcheney, Asilomarlyonsrajcheney, CGNetTSP} similarly decompose $\bm{c}$ via (\ref{eqn:CG}), but restrict $\bm{z} = h(\bm{x})$ where $\bm{x}\sim\mathcal{N}(\bm{0},I)$ and $h$ is an invertible nonlinear function. For this choice of $\bm{z}$, $\mathcal{R}$ in (\ref{eqn:cost function}) is set at $\mathcal{R}(\bm{z}) = \mu\norm{h^{-1}(\bm{z})}_2^2$, for scalar $\mu> 0$, to enforce normality of $\bm{x}$. Furthermore, previous work~\cite{APSIPAlyonsrajcheney, Asilomarlyonsrajcheney, CGNetTSP} restricts $P_u$ in (\ref{eqn:cost function}) to be a scaled identity matrix. Our G-CG-LS algorithm naturally generalizes these prior CG-based iterative algorithm cost functions to a cost function with implicit scale regularization, $\mathcal{R}$, and general covariance matrix structure. By applying algorithm unrolling to G-CG-LS, as shown in Section~\ref{sec:DR-CG-Net}, $\mathcal{R}$ and $P_u$ can be optimally learned to produce a DNN with greater representational capacity and applicability than the unrolled CG-based DNNs proposed in~\cite{APSIPAlyonsrajcheney, Asilomarlyonsrajcheney, CGNetTSP}.

\subsection{Scale Variable Update Methods} \label{sec:scale variable update methods}
Two methods we employ to minimize (\ref{eqn:cost function}) w.r.t. $\bm{z}$ are detailed below where $\eta > 0$ is a real-valued step size. Note that, for both methods, we define $A_{\bm{u}} = A\textnormal{Diag}(\bm{u})$.
\paragraph{Projected Gradient Descent (PGD)}
\begin{align*}
    g(\bm{z},\bm{u}) &\coloneqq \mathcal{P}_{\mathfrak{Z}}\left(\bm{z} - \eta\nabla_{\bm{z}} F(\bm{u},\bm{z})\right) \\
    &= \mathcal{P}_{\mathfrak{Z}}\left(\bm{z}-\eta\left(A_{\bm{u}}^T(A_{\bm{u}}\bm{z} - \bm{y}) + \nabla \mathcal{R}(\bm{z})\right)\right)
\end{align*}
where $\mathcal{P}_{\mathfrak{Z}}$ is the projection onto $\mathfrak{Z}.$
\paragraph{Iterative Shrinkage and Thresholding (ISTA)}
\begin{align*}
     g(\bm{z},\bm{u}) &\coloneqq \textnormal{prox}_{\eta\mathcal{R}}(\bm{z}-\eta A_{\bm{u}}^T(A_{\bm{u}}\bm{z} - \bm{y})) \\
    &= \underset{\bm{\zeta}\in \mathfrak{Z}}{\arg\min}\,\, \frac{1}{2}\norm{\bm{\zeta} - (\bm{z}-\eta A_{\bm{u}}^T(A_{\bm{u}}\bm{z} - \bm{y}))}_2^2 + \eta\mathcal{R}(\bm{\zeta})
\end{align*}
where prox$_{f}$ is the proximal operator of $f$ and is well-defined for convex $f$. We remark that for a non-smooth, convex function, $f$, the proximal operator is an optimization tool as fixed points of $\textnormal{prox}_f$ minimize $f$. The general ISTA method, above in \textit{(b)}, which is equivalent to proximal gradient descent, is an optimization method for the sum of a convex, differentiable function and a convex, non-smooth function~\cite{fast_ISTA}. Using the proximal operator on the non-smooth function, fixed points of ISTA are optimality points of the original function sum (see Appendix~\ref{apndx:ISTA bounds}).

The PGD and ISTA methods were chosen for their simplicity of implementation, provable convergence properties, and the success of unrolling ISTA for linear inverse problems~\cite{MADUN, zhang2018ista, learned_ISTA, xiang2021FISTANet}. Additional details about PGD and ISTA, including backtracking linesearch methods and descent bounds, are given in Appendices~\ref{apndx:PGD bounds} and~\ref{apndx:ISTA bounds}, respectively.

\subsection{Convergence of G-CG-LS}

This subsection provides theoretical convergence properties of G-CG-LS and can be skipped by those focused on implementation details. Convergence is proven in the Appendix under certain combinations of the following assumptions:
\begin{itemize}
    \item[$(\mathcal{A}1)$] $\mathcal{R}$ is bounded below and satisfies $\lim_{z_i\to\infty} \mathcal{R}(\bm{z})\to\infty$ for all $i = 1, 2, \ldots, n$.
    \item[$(\mathcal{A}2)$] $\mathcal{R}$ is convex but possibly non-smooth.
    \item[$(\mathcal{A}3)$] $\mathcal{R}$ is twice continuously differentiable (i.e. $\mathcal{R}\in C^2$). 
\end{itemize}

We first remark that (\ref{eqn:cost function}) is strongly convex w.r.t. $\bm{u}$, with strength controlled by the spectral radius of $P_u$~\cite{bertero1988linear}, and thus (\ref{eqn:cost function}) has no maxima.

\begin{proposition}\label{prop:convergence of const function values}
Let $\mathcal{R}$ satisfy $(\mathcal{A}1)$. If $\mathcal{R}$ satisfies $(\mathcal{A}2)$ for ISTA G-CG-LS or $(\mathcal{A}3)$ for PGD G-CG-LS, then the sequence $\{F(\bm{u}_k,\bm{z}_k)\}_{k = 1}^\infty$ converges.
\end{proposition}

Using Proposition~\ref{prop:convergence of const function values}, G-CG-LS converges to stationary points of (\ref{eqn:cost function}) defined in Definition~\ref{def:optimality conditions} of the Appendix.

\begin{theorem} \label{thm:ISTA and PGD convergence}
Let $\mathcal{R}$ satisfy $(\mathcal{A}1)$. If $\mathcal{R}$ satisfies $(\mathcal{A}2)$ for ISTA G-CG-LS or $(\mathcal{A}3)$ for PGD G-CG-LS, then every limit point of the sequence $\{(\bm{u}_k,\bm{z}_k)\}_{k = 1}^\infty$ is a stationary point of (\ref{eqn:cost function}).
\end{theorem}

Proof of Proposition~\ref{prop:convergence of const function values} and Theorem~\ref{thm:ISTA and PGD convergence} are in Appendix~\ref{apndx:ISTA and GD cost function convergence} and Appendix~\ref{apndx:ISTA and GD estimate convergence}, respectively. The proof of Theorem~\ref{thm:ISTA and PGD convergence} is inspired by ideas from \cite{bertsekas2016nonlinearprogramming, fast_ISTA, xu2013block}, which consider the convergence of PGD and ISTA-type updates. Where \cite{bertsekas2016nonlinearprogramming, fast_ISTA} consider a single block update, which we can transfer to individual updates of $\bm{z}$ when $\bm{u}$ is fixed, we distinctly have two block updates defining each iteration. While \cite{xu2013block} considers multiblock updates, each block is updated by a single step of an identical method. Instead, we consider multiple steps and unique methods defining each block update.

Next, we discuss the convergence of the G-CG-LS estimates. For this, we say a sequence converges to set $\mathcal{S}$ if the sequence values become arbitrarily close to elements of $\mathcal{S}$.

\begin{proposition}\label{prop:convergence to set of const value}
Let the assumptions of Theorem~\ref{thm:ISTA and PGD convergence} hold. Then the sequence $\{(\bm{u}_k,\bm{z}_k)\}_{k = 1}^\infty$ generated by ISTA G-CG-LS or PGD G-CG-LS converges to a closed connected set of stationary points of constant cost function value.
\end{proposition}

\begin{corollary}\label{corollary:convergence}
Let the assumptions of Theorem~\ref{thm:ISTA and PGD convergence} hold. Additionally, if $\mathcal{R}$ is chosen such that (\ref{eqn:cost function}) has nondegenerate stationary points then the ISTA G-CG-LS and PGD G-CG-LS sequences $\{(\bm{u}_k,\bm{z}_k)\}_{k = 1}^\infty$ converge.
\end{corollary}

Proofs of Proposition~\ref{prop:convergence to set of const value} and Corollary~\ref{corollary:convergence} are found in Appendix~\ref{apndx:covergence to set of const value}. We remark in choosing $\mathcal{R}(\bm{z})$ to be positively defined on an open domain, e.g. $\mathcal{R}(\bm{z}) = \norm{\log(\bm{z})}_2^2$ defined on $(0,\infty)^n$, all stationary points are internal points and, consequently, the assumptions of Theorem~\ref{thm:ISTA and PGD convergence} and Corollary~\ref{corollary:convergence} hold, implying G-CG-LS converges to a limit point of zero gradient (or zero subdifferential when $\mathcal{R}$ is convex). In choosing $\mathcal{R}(\bm{z})\propto -\log(p_{\bm{z}}(\bm{z}))$, as in a MAP estimate, many prior distributions, $p_{\bm{z}}$, produce $\mathcal{R}(\bm{z})$ defined only on the open domain $(0,\infty)^n$. Specifically, any twice continuously differentiable distribution satisfying $\lim_{z_i\to 0} p_{\bm{z}}(\bm{z})\to 0$ will produce $\mathcal{R}$  only defined on $(0,\infty)^n$, thus satisfying the PGD assumptions of Theorem~\ref{thm:ISTA and PGD convergence}. Examples of such $\bm{z}$ distributions useful in modeling image coefficients include log-normal and Gamma$(\alpha, 1)$ for $\alpha > 1$ distributions \cite{Wavelet_Trees, Scale_Mixtures, HB-MAP}.

\section{DR-CG-Net}\label{sec:DR-CG-Net}

   %%%%%%%%%%%%%%%%%%%%%%%%% DR-CG-Net Diagram %%%%%%%%%%%%%%%%%%%%%%%%%%%
\begin{figure*}[!t]
    \centering
    \begin{subfigure}{\textwidth}
    \centering
    \includegraphics[scale = 0.65]{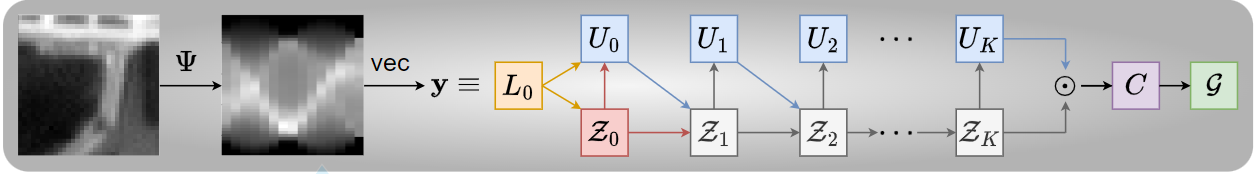}
    \caption{End-to-end network structure of DR-CG-Net.}
    \label{fig:DR-CG-Net}
\end{subfigure}

\vspace*{0.15cm}

\begin{subfigure}{0.38\textwidth}
    \centering
    \includegraphics[scale = 0.65]{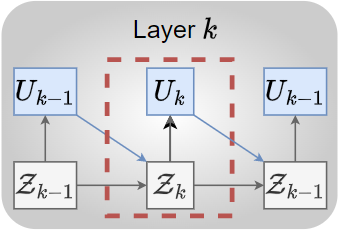}
    \caption{Layer $k$ analogous to iteration $k$ of Algorithm~\ref{alg:CG-LS}.}
    \label{fig:DR-CG-Net block}
\end{subfigure}
\begin{subfigure}{0.61\textwidth}
    \centering
    \includegraphics[scale = 0.65]{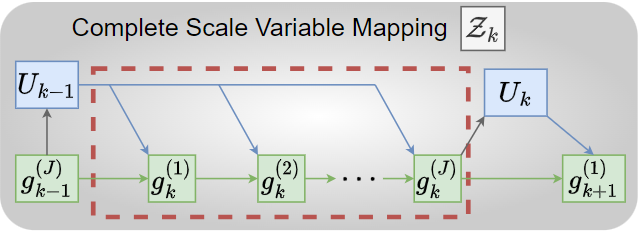}
    \caption{Complete scale variable mapping, $\mathcal{Z}_k$, producing estimate $\bm{z}_k$ in Algorithm~\ref{alg:CG-LS}.}
    \label{fig:DR-CG-Net scale mapping module}
\end{subfigure}

\vspace*{0.15cm}

\begin{subfigure}{\textwidth}
    \centering
    \begin{subfigure}{0.49\textwidth}
    \centering
    \includegraphics[scale = 0.75]{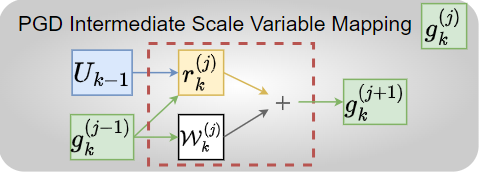}
\end{subfigure}
\begin{subfigure}{0.49\textwidth}
    \centering
    \includegraphics[scale = 0.75]{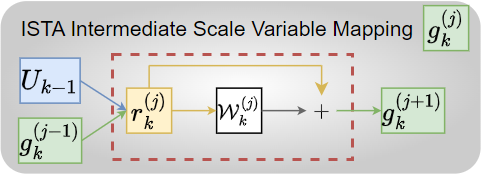}  
\end{subfigure}
    \caption{Intermediate scale variable mapping, $g_k^{(j)}$, for PGD (left) and ISTA (right) which produces the estimate, $\bm{z}_k^{(j)}$, in Algorithm~\ref{alg:CG-LS}.}
    \label{fig:DR-CG-Net z update Module}
\end{subfigure}

\vspace*{0.15cm}

\begin{subfigure}{\textwidth}
    \centering
    \includegraphics[scale = 0.85]{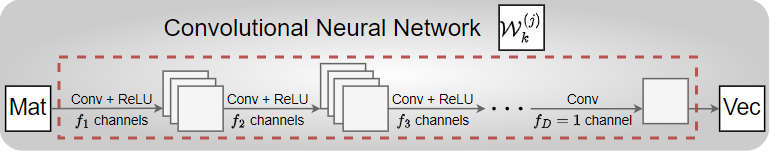}
    \caption{Network representing $\nabla \mathcal{R}$ or prox$_{\eta\mathcal{R}}$ for PGD or ISTA, respectively. No bias is added, ReLU activation functions are employed, and $f_1, f_2, \ldots,$ and $f_D$ denote the integer number of output filter channels. \textit{Mat} converts a vector into a matrix and \textit{Vec} inverts this process.}
    \label{fig:DR-CG-Net CNN}
\end{subfigure}
\caption{End-to-end network structure for DR-CG-Net, the unrolled deep neural network of Algorithm~\ref{alg:CG-LS}, is shown in (\ref{fig:DR-CG-Net}). DR-CG-Net consists of an input block, $L_0$, initialization block, $Z_0$, $K+1$ Tikhonov blocks, $U_k$, $K$ complete scale variable mappings, $\mathcal{Z}_k$, a Hadamard product block, $C$, and an optional refinement block, $\mathcal{G}$. Each $\mathcal{Z}_k$, with structure in (\ref{fig:DR-CG-Net scale mapping module}), consists of $J$ updates $g_k^{(j)}$ further detailed in (\ref{fig:DR-CG-Net z update Module}). Each $g_k^{(k)}$ consists of a data fidelity gradient descent step, $r_k^{(j)}$, added to a convolutional neural network, $\mathcal{W}_k^{(j)}$ in (\ref{fig:DR-CG-Net CNN}), and corresponds to an intermediate update of the $\bm{z}$ variable.}
\label{fig:DR-CG-Net structure}
\end{figure*}
    %%%%%%%%%%%%%%%%%%%%%%%%%%%%%%%%%%%%%%%%%%%%%%%%%%

\subsection{Network Structure}
We create a DNN named DR-CG-Net, with the end-to-end structure shown in Fig.~\ref{fig:DR-CG-Net}, by applying algorithm unrolling to G-CG-LS in Algorithm~\ref{alg:CG-LS}. Instead of requiring a specified $\mathcal{R},$ DR-CG-Net learns $\mathcal{R}$ through a subnetwork representing either $\nabla \mathcal{R}$ or prox$_{\eta\mathcal{R}}$.

Let $\mathcal{V}_k^{(j)}:\mathbb{R}^n\to\mathbb{R}^n$, for $k = 1, 2, \ldots, K$ and $j = 1, 2, \ldots, J$, be a subnetwork. That is, each $\mathcal{V}_k^{(j)}$ is a collection of layers mapping from $\mathbb{R}^n$ to $\mathbb{R}^n$. 
The particular form of $\mathcal{V}_k^{(j)}$ we employ is given in Section \ref{sec:DR-CG-Net params}.
Following the scale variable updates of Section~\ref{sec:scale variable update methods} we define the intermediate scale variable mapping, $g_k^{(j)}$, as
\begin{align}
    g_k^{(j)}(\bm{z},\bm{u}) \coloneqq \begin{cases}
   \textnormal{ReLU}\left(r_k^{(j)}(\bm{z},\bm{u}) + \mathcal{V}_k^{(j)}(\bm{z})\right) & \textnormal{ PGD} \\
   \textnormal{ReLU}\left(\mathcal{V}_k^{(j)}\left(r_k^{(j)}(\bm{z},\bm{u})\right)\right) & \textnormal{ ISTA}
    \end{cases} \label{eqn:intermediate mapping module}
\end{align}
where $r_k^{(j)}$ is a data fidelity gradient descent step of (\ref{eqn:cost function}) in $\bm{z}$,
\begin{align}
    r_k^{(j)}(\bm{z},\bm{u}) = \bm{z} - \eta_k^{(j)}A_{\bm{u}}^T(A_{\bm{u}}\bm{z}-\bm{y}) \label{eqn:data fidelity gradient}
\end{align}
for a step size $\eta_k^{(j)}$. Note that we apply the ReLU function componentwise in (\ref{eqn:intermediate mapping module}), where ReLU $\equiv \mathcal{P}_{[0,\infty)}$, to ensure all $\bm{z}$ estimates in DR-CG-Net maintain positive entries.

Each layer $k$ of DR-CG-Net,  shown in Fig.~\ref{fig:DR-CG-Net block}, is broken down into a complete scale variable mapping block, $\mathcal{Z}_k$ shown in Fig.~\ref{fig:DR-CG-Net scale mapping module}, and a Tikhonov solution block, $U_k$, so that layer $k$ corresponds to iteration $k$ of Algorithm~\ref{alg:CG-LS}. As in Algorithm~\ref{alg:CG-LS}, $\mathcal{Z}_k$ consists of a composition of the $J$ scale variable update blocks $g_k^{(1)},\ldots, g_k^{(J)}$ shown in Fig.~\ref{fig:DR-CG-Net z update Module} for PGD or ISTA.

Mathematically detailing the DR-CG-Net blocks we have:

\noindent\begin{tikzpicture}
\node[scale = .9, fill = peachfill, draw = darkorangeborder, thick] at (0, 0) {$\phantom{L}$};
\node[scale = 1] at (0, 0) {$L_0$};
\node[anchor = west] at (0.15,0) {$= \bm{y}$ is the input measurements to the network.};
\end{tikzpicture} 

\vspace*{-.1cm}

\noindent\begin{tikzpicture}
\node[scale = .9, fill = pinkfill, draw = darkredborder, thick] at (0, 0) {$\phantom{Z}$};
\node[scale = 1] at (0, 0) {$\mathcal{Z}_0$};
\node[anchor = west] at (0.15,0) {$=\mathcal{P}_{[0,b]^n}(\hat{A}^T\bm{y})$, for $\hat{A} = A/||A||_2$, is the initial};
\node[anchor = west] at (0.15,-.45) {estimate of $\bm{z}$ from line~\ref{line:initial} of Algorithm~\ref{alg:CG-LS}.};
\end{tikzpicture} 

\vspace*{-.15cm}

\noindent\begin{tikzpicture}
\node[scale = .9, fill = bluefill, draw = blueborder, thick] at (0, 0) {$\phantom{U}$};
\node[scale = 1] at (0, 0) {$U_k$};
\node[anchor = west] at (0.15,0) {$ = \mathcal{T}(g_k^{(J)}; P_k)$ (or $\widetilde{\mathcal{T}}(U_{k-1}, g_k^{(J)}; P_k)$ for NAGD), for};
\node[anchor = west] at (0.15,-.45) {covariance matrix $P_k$, is the Tikhonov estimate corresp-};
\node[anchor = west] at (0.15,-.9) {onding to line~\ref{line:initial} and~\ref{line:u update} of Algorithm~\ref{alg:CG-LS}.};
\end{tikzpicture}

\vspace*{-.15cm}

\noindent\begin{tikzpicture}
\node[anchor = west] at (0,0) {The $k$th complete scale variable mapping};
\node[scale = 0.9, fill = palegray, draw = black] at (6.5, 0) {$\phantom{Z}$};
\node[scale = 1] at (6.5, 0) {$\mathcal{Z}_k$};
\node[anchor = west] at (6.8,0) {contains:};
\end{tikzpicture}

\hspace*{-.16cm}\begin{tikzpicture}
\node[scale = 1.2, fill = greenfill, draw = greenborder, thick] at (0, 0) {$\phantom{Z}$};
\node[scale = 1] at (0, 0) {$g_k^{(j)}$};
\node[anchor = west] at (0.2,0) {\scalebox{.95}{$= g_k^{(j)}(g_k^{(j-1)},U_{k-1})$} is an intermediate scale variable};
\node[anchor = west] at (0.2,-.45) {mapping analogous to $\bm{z}_k^{(j)}$ on line~\ref{line:z update} in Algorithm~\ref{alg:CG-LS}.};
\end{tikzpicture}

\hspace*{-.16cm}\begin{tikzpicture}
\node[scale = 1.2, fill = yellowfill, draw = yellowborder, thick] at (0, 0) {$\phantom{Z}$};
\node[scale = 1] at (0, 0) {$r_k^{(j)}$};
\node[anchor = west] at (0.2,0) {$= r_k^{(j)}(g_k^{(j-1)}, U_{k-1})$ is the data fidelity gradient step.};
\end{tikzpicture}

\hspace*{-.16cm}\begin{tikzpicture}
\node[scale = 1.2, fill = palegray, draw = black] at (0, 0) {$\phantom{Z_k}$};
\node[scale = 1] at (0, 0) {$\mathcal{W}_k^{(j)}$};
\node[anchor = west] at (0.4,0) {$= D$ layer convolutional neural network.};
\end{tikzpicture}

\noindent\begin{tikzpicture}
\node[scale = .9, fill = purplefill, draw = purpleborder, thick] at (0, 0) {$\phantom{C}$};
\node[scale = 1] at (0, 0) {$C$};
\node[anchor = west] at (0.15,0) {$= U_K\odot \mathcal{Z}_K$ is an estimate of the signal, $\bm{c}$.};
\end{tikzpicture}

\noindent\begin{tikzpicture}
\node[scale = 1, fill = greenfill, draw = greenborder, thick] at (0, 0) {$\phantom{Z}$};
\node[scale = 1] at (0, 0) {$\mathcal{G}$};
\node[anchor = west] at (0.2,0) {\scalebox{.95}{$= g_{K+1}^{(1)}(C,\bm{1})$} is an optional refinement of the estimated};
\node[anchor = west] at (0.2,-.45) {signal, $\bm{c}$, producing the DR-CG-Net output.};
\end{tikzpicture}

Note, to simplify notation, we let $g_k^{(0)} = g_{k-1}^{(J)}$, $U_{-1} = \bm{0}$, and $\bm{1} \in\mathbb{R}^n$ be a vector of ones. We remark that the optional refinement block $\mathcal{G}$ deviates from the structure of G-CG-LS and instead corresponds to an unrolled PGD or ISTA step on the cost function $\widetilde{F}(\bm{c}) \coloneqq \frac{1}{2}\norm{\bm{y} - A\bm{c}}_2^2 + \widetilde{\mathcal{R}}(\bm{c})$ where $\widetilde{\mathcal{R}}(\bm{c})$ is a learned implicit regularization on the total signal, $\bm{c}$. 
As (\ref{eqn:cost function}) is a special case of $\widetilde{F}$, where the CG decomposition from (\ref{eqn:CG}) is used to split $\bm{c}$, the refinement block simultaneously updates both $\bm{z}$ and $\bm{u}$ such that the interrelationship between these fields is best exploited for the signal of interest. 

The refinement block is motivated, in part, by prior art approaches, such as MADUN~\cite{MADUN} and ISTA-Net$^+$~\cite{zhang2018ista}, that model each DNN layer as an ISTA-type step on the cost function $\widetilde{F}$. As such, our DR-CG-Net approach first exploits the powerful CG prior through minimizing (\ref{eqn:cost function}) to obtain a high quality estimate for the inverse problem to (\ref{eqn:linear_msrmt}), and subsequently fine-tunes this estimate through the optional refinement block that mimics a single layer of these prior art methods. Empirically shown in Section~\ref{sec:refinement block}, the unrolled G-CG-LS portion of DR-CG-Net, that is up to and including block $C$, is the core component in signal reconstruction performance and that the refinement block marginally refines, e.g. denoises, the final estimate $C$.

\begin{figure}[!t]
    \centering
    \includegraphics[width=\columnwidth]{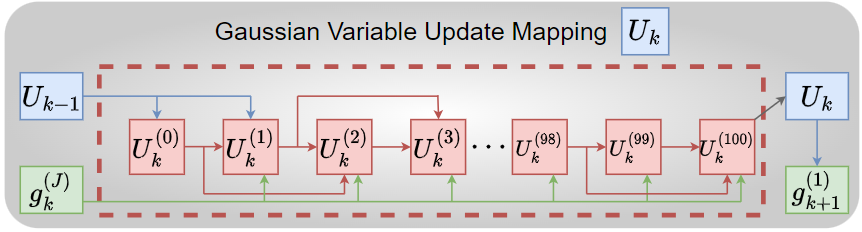}
    \caption{Structure of the $\bm{u}$ update block, $U_k$, for using gradient descent steps with Nesterov momentum.}
    \label{fig:DR-CG-Net u GD block}
\end{figure}

Finally, as discussed in Section~\ref{sec:interative_alg}, for significantly high-dimensional signals of interest, the $U_k$ block contains $J_u$ NAGD steps as detailed in Fig.~\ref{fig:DR-CG-Net u GD block}. Using (\ref{eqn:G-CG-LS GD u step}), we define the intermediate Gaussian variable mapping, $g_u(\bm{u}, \bm{w}, \bm{z}; \beta)$, as
\begin{align*}
    g_u(\bm{u}, \bm{w}, \bm{z}; \beta) \coloneqq r_u(\bm{u},\bm{z}) + \beta \left(r_u(\bm{u},\bm{z}) - r_u(\bm{w},\bm{z})\right).
\end{align*}
Then, $U_k^{(j)}$ in Fig.~\ref{fig:DR-CG-Net u GD block} is given by

\noindent\begin{tikzpicture}
\node[scale = .85, fill = pinkfill, draw = darkredborder, thick] at (0, 0) {$\phantom{U_k^{(j)}}$};
\node[scale = .9] at (0, 0) {$U_k^{(j)}$};
\node[anchor = west] at (0.4,0) {\scalebox{0.97}{$=g_u\left(U_k^{(j-1)}, U_k^{(j-2)}, g_k^{(J)}; 1-\frac{3}{6+j}\right)$} is a NAGD step.};
\end{tikzpicture}
Note that we set $U_k^{(0)} = U_k^{(-1)} \equiv U_{k-1}$ and choose an adaptive momentum parameter of $1-\frac{3}{6+j}$, for $j$ the NAGD step count, as proposed in~\cite{nesterov1983method}. For this setting, the $U_k$ block is given by the approximation of the Tikhonov solution $\widetilde{\mathcal{T}}$ rather the exact Tikhonov solution $\mathcal{T}$.

%%%%%%%%%%%%%%%%%%%%%%%%% DR-CG-Net GD u Diagram %%%%%%%%%%%%%%%%%%%%%%%%%%%
%%%%%%%%%%%%%%%%%%%%%%%%%%%%%%%%%%%%%%%%%%%%%%%%%%%%%%%%%%

\subsection{Network Parameters and Subnetwork} \label{sec:DR-CG-Net params}

For every $k = 0, 1, \ldots, K$, the layer $U_k$ is parameterized by a covariance matrix, $P_k$. To reduce the number of parameters learned by the network and for consistency in $P_k$ representing the covariance matrix of $\bm{u}$, DR-CG-Net learns a single covariance matrix $P$ and constrains $P_1 = \cdots = P_K = P.$ Furthermore, we consider the possibility of a structured covariance matrix where $P$ is either a scaled identity, diagonal, tridiagonal, or full matrix. Imposing a covariance structure may be desirable or advantageous, for example, to ensure only local reinforcement.  

We remark that to ensure $P$ is a covariance matrix, i.e. symmetric and positive definite, we impose, for $\epsilon > 0$ a small fixed real number, one of the following structures
\begin{align*}
    P = \begin{cases}
        \max\{\lambda, \epsilon\} I & \textnormal{ Scaled Identity} \\
        \textnormal{Diag}([\max\{\lambda_i, \epsilon\}]_{i =1}^n) & \textnormal{ Diagonal} \\
        L_{\textnormal{tri}}L_{\textnormal{tri}}^T + \epsilon I & \textnormal{ Tridiagonal} \\
        LL^T + \epsilon I & \textnormal{ Full}.
    \end{cases}
\end{align*}
In the scaled identity case, only a constant $\lambda$ is learned. In the diagonal case, a vector $\bm{\lambda} = [\lambda_i]_{i = 1}^n$ is learned. In the tridiagonal case, two vectors $\bm{\lambda}_1\in\mathbb{R}^n$ and $\bm{\lambda}_2\in\mathbb{R}^{n-1}$ are learned such that the lower triangular matrix component $L_{\textnormal{tri}}$ is formed by placing $\bm{\lambda}_1$ on the diagonal and $\bm{\lambda}_2$ on the first subdiagonal. Finally, in the case of a full covariance matrix, an entire lower triangular matrix, $L$, is learned.

Next, each $\mathcal{Z}_k$ is parameterized both by a collection of step sizes $\{\eta_k^{(1)}, \ldots, \eta_k^{(J)}\}$ (as $\eta_k^{(j)}$ parameterizes the data fidelity gradient step, $r_k^{(j)}$) and by the parameters of the subnetworks $\mathcal{V}_k^{(1)}, \ldots, \mathcal{V}_k^{(J)}$. Similarly, the refinement block is parameterized by a step size $\eta_{K+1}^{(1)}$ and subnetwork $\mathcal{V}_{K+1}^{(1)}.$

For the step sizes, we take $\eta_k^{(j)}$ from (\ref{eqn:data fidelity gradient}) as
\begin{align*}
    \eta_k^{(j)} = \delta_k^{(j)}\begin{cases}
        1 & \norm{A_{\bm{u}}^T(A_{\bm{u}}\bm{z}-\bm{y})}_2 \leq \gamma_{\max} \\
        \frac{\gamma_{\max}}{\norm{A_{\bm{u}}^T(A_{\bm{u}}\bm{z}-\bm{y})}_2} & \textnormal{ else}
    \end{cases}
\end{align*}
for fixed constant $\gamma_{\max} > 0$ and a learned real-valued parameter $\delta_k^{(j)}$. This is a slight variation on normalized gradient descent~\cite{murray2019normalizedgradientdescent} for the data fidelity term of (\ref{eqn:cost function}), which we empirically find to provide stability. In particular, normalized gradient descent supplies sample-specific step size adjustments to prevent a large and counterproductive gradient descent step.

For the subnetworks, let $\mathcal{W}_k^{(j)}$ be a CNN of depth $D$ using ReLU activation functions where the convolutions in layer $d$ use $k_d\times k_d$ kernel size and produce $f_d$ output filter channels as shown in Fig.~\ref{fig:DR-CG-Net CNN}. Note, zero padding is applied to each filter channel of the input such that that the output at each filter channel is the same size as the input. Furthermore, we take $f_D = 1$ such that for $X\in\mathbb{R}^{n\times n}$ we have $\mathcal{W}_k^{(j)}(X)\in\mathbb{R}^{n\times n}.$ Next, for $\bm{x}\in\mathbb{R}^{n^2}$ we define mat$(\bm{x})$ to be $\bm{x}$ reshaped into a $n\times n$ matrix. Similarly, for $X\in\mathbb{R}^{n\times n}$, we define vec$(X)$ to be $X$ reshaped into a vector of size $n^2$ such that vec(mat($\bm{x})) = \bm{x}$. Then we set
\begin{align*}
    \mathcal{V}_k^{(j)}(\bm{x}) = \begin{cases}
        \textnormal{vec}(\mathcal{W}_k^{(j)}(\textnormal{mat}(\bm{x}))) & \textnormal{PGD} \\
        \bm{x} + \textnormal{vec}(\mathcal{W}_k^{(j)}(\textnormal{mat}(\bm{x}))) & \textnormal{ISTA}.
    \end{cases}
\end{align*}
Define $W_{k,d}^{(j)}$, for $d = 1, 2, \ldots, D$, as the convolutional weight kernels of $\mathcal{W}_k^{(j)}$, which also parameterize the subnetwork $\mathcal{V}_k^{(j)}$.

From the above, the DR-CG-Net parameters are 
\[
\bm{\Theta} = \{P\}\cup\left\{\delta_k^{(j)}, W_{k,d}^{(j)}, \delta_{K+1}^{(1)}, W_{K+1, d}^{(1)}\right\}\middlescript{\substack{j = 1, 2, \ldots, J \\ k = 1, 2, \ldots, K \\ d = 1, 2, \ldots, D}}.
\]
Let $f_0 = 1$, $p = \sum_{d = 1}^D f_{d-1}f_d k_d^2$, and $\dim(\mathcal{P}) =1, n, 2n-1,$ or $n(n+1)/2$ for the scaled identity, diagonal, tridiagonal, or full covariance matrix structures, respectively. Then DR-CG-Net has $\dim(\mathcal{P})+(KJ+1)(p+1)$ total parameters.

\subsection{Loss Function}

The DR-CG-Net parameters are trained by minimizing the MAE loss function given as
\begin{align}
    \mathcal{L}_{\mathcal{B}}(\bm{\Theta}) = \frac{1}{|\mathcal{B}|}\frac{1}{n}\sum_{(\overline{\bm{y}}_i, \overline{\bm{c}}_i)\in \mathcal{B}} \norm{\widehat{\bm{c}}(\overline{\bm{y}}_i; \bm{\Theta}) - \overline{\bm{c}}_i}_1 \label{eqn:loss}
\end{align}
where $\mathcal{B}$ is a batch of data points.
The MAE loss function is optimized with adaptive moment estimation (Adam)~\cite{ADAM}, which is a stochastic, gradient-based optimizer. The gradient, $\nabla_{\bm{\Theta}} \mathcal{L}_{\mathcal{B}}$, for Adam is calculated via backpropagation through the network, which we implement with automatic differentiation using TensorFlow~\cite{abadi2016tensorflow}.

\section{Numerical Results}

We consider two types of DR-CG-Nets, called PGD DR-CG-Net and ISTA DR-CG-Net, corresponding, respectively, to using the PGD or ISTA intermediate scale variable mapping given in (\ref{eqn:intermediate mapping module}). For unrolled iterations we set $(K, J) = (3, 4)$ for $32\times 32$ images and $(K, J) = (1, 24)$ for all larger images. Each $\mathcal{V}_k^{(j)}$ is a CNN with depth $D = 8$. The network size was chosen empirically such that the time to complete a signal reconstruction was reasonably quick while still producing excellent reconstructions on a validation dataset. Every convolution uses a $3\times 3$ kernel initialized according to the Glorot Uniform distribution~\cite{CNN_initialization} with ReLU activation functions and $f_1 = \cdots = f_7 = 32$ and $f_8 = 1$ filter channels.

We initialize each step size as $\delta_k^{(j)} = 1$, set $\gamma_{\max} = 1$, and fix $\epsilon = 10^{-4}.$ Each $P_u$ structure is initialized as a diagonal matrix with $0.1$ or $10$ on the diagonal for Radon transform or Gaussian measurements, respectively. Additionally, for the initial $\bm{z}$ estimate, we set $\mathcal{P}_{[0,b]^n} = \mathcal{P}_{[0, 10]^n}.$  We train all DR-CG-Nets for 2000 epochs at a learning rate of $10^{-4}$ implementing early stopping to prevent overfitting.

We compare DR-CG-Net against ten state-of-the-art methods: (i) compound Gaussian network (CG-Net)~\cite{CGNetTSP, APSIPAlyonsrajcheney}, (ii) memory augmented deep unfolding network (MADUN)~\cite{MADUN}, (iii) ISTA-Net$^+$~\cite{zhang2018ista}, (iv) FISTA-Net~\cite{xiang2021FISTANet}, (v) iPiano-Net~\cite{su2020iPianoNet}, (vi) ReconNet~\cite{reconnet}, (vii) LEARN$^{++}$~\cite{zhang2022learn++}, (viii) Learned Primal-Dual (LPD)~\cite{adler2018LPD}, (ix) FBPConvNet~\cite{jin2017FBPConvNet}, and (x) iRadonMAP~\cite{he2020iRadonMAP}. Although memory augmented proximal unrolled network (MAPUN)~\cite{song2023MAPUN} was considered, due to the similarity in performance to MADUN only MADUN results are shown. Finally, two model-based approaches, filtered backprojection (FBP)~\cite{radontransform} and super-voxel model-based iterative reconstruction (svMBIR)~\cite{svmbir}, are compared. As these model-based approaches are generally not as competitive as the top deep-learning methods, their results are relegated to the single plot of Fig.~\ref{fig:reconstructions64} to provide a baseline.

Note that LEARN$^{++}$, LPD, FBPConvNet, and iRadonMAP are CT-specific methods relying on the CT sinogram measurement structure. On the other hand, MADUN, ISTA-Net$^+$, FISTA-Net, iPiano-Net, and ReconNet are reconstruction methods with particular application in image CS. Furthermore, we remark that CG-Net, MADUN, ISTA-Net$^+$, FISTA-Net, iPiano-Net, LEARN$^{++}$, and LPD are DNNs formed by algorithm unrolling while ReconNet, iRadonMAP, and FBPConvNet are instead standard DNNs. Lastly, as with DR-CG-Net, a refinement block is added to, and trained with, each CG-Net.

\begin{figure*}[!t]
\centering
\begin{subfigure}{0.32\textwidth}
    \centering
    \includegraphics[width=\textwidth]{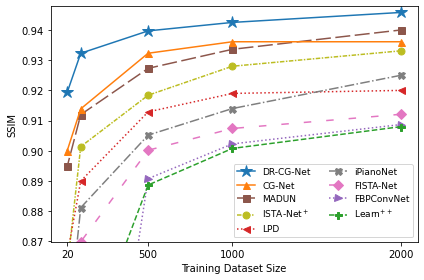}
    \caption{Fifteen uniform angles and 60dB SNR.}
    \label{fig:compare_15_60}
\end{subfigure}
\begin{subfigure}{0.32\textwidth}
    \centering
    \includegraphics[width=\textwidth]{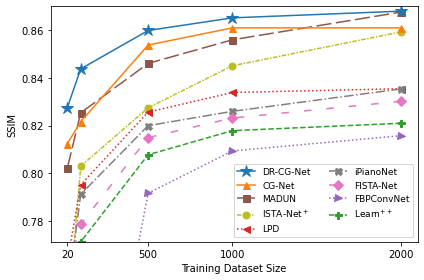}
    \caption{Ten uniform angles and 60dB SNR.}
    \label{fig:compare_10_60}
\end{subfigure}
\begin{subfigure}{0.32\textwidth}
    \centering
    \includegraphics[width=\textwidth]{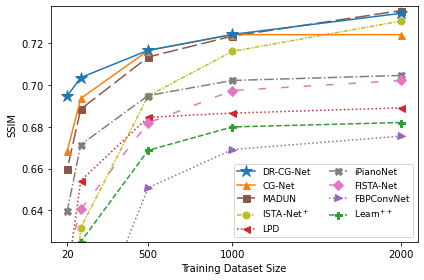}
    \caption{Six uniform angles and 60dB SNR.}
    \label{fig:compare_6_60}
\end{subfigure}
\begin{subfigure}{0.32\textwidth}
    \centering
    \includegraphics[width=\textwidth]{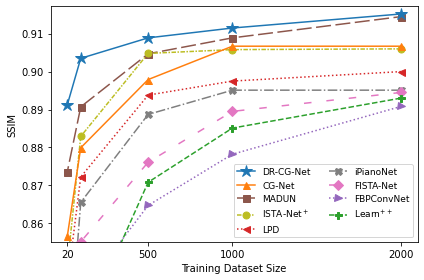}
    \caption{Fifteen uniform angles and 40dB SNR.}
    \label{fig:compare_15_40}
\end{subfigure}
\begin{subfigure}{0.32\textwidth}
    \centering
    \includegraphics[width=\textwidth]{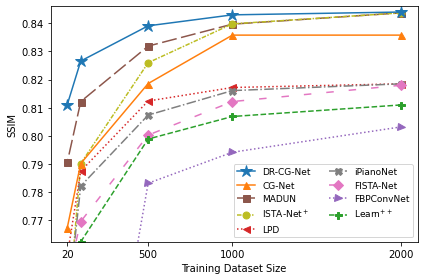}
    \caption{Ten uniform angles and 40dB SNR.}
    \label{fig:compare_10_40}
\end{subfigure}
\begin{subfigure}{0.32\textwidth}
    \centering
    \includegraphics[width=\textwidth]{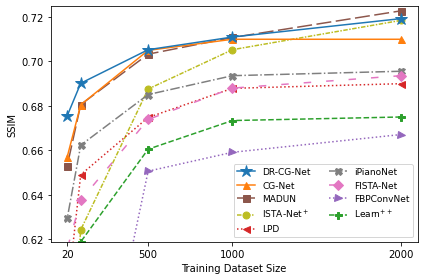}
    \caption{Six uniform angles and 40dB SNR.}
    \label{fig:compare_6_40}
\end{subfigure}

\caption{Average test image reconstruction SSIM when varying the amount of CIFAR10 data in training nine machine learning-based image reconstruction methods. Here, the sensing matrices, $\Psi$, are a Radon transform at 15, 10, or 6 uniformly spaced angles, $\Phi = I$, and the measurement SNR is 60dB or 40dB. \textbf{Our DR-CG-Net method outperforms the compared prior art methods, in all scenarios, and does so appreciably in low training}.}
\label{fig:compare}
\end{figure*}

\subsection{Data} \label{sec:data}

We use $32\times 32$ CIFAR10 images~\cite{CIFAR10}, $64\times 64$ CalTech101 images~\cite{CalTech101}, and $128\times 128$ LoDoPaB-CT images~\cite{lodopabct}. Each image has been converted to a single-channel grayscale image, scaled down by the maximum pixel value, and vectorized. We apply a given sensing method, $\Psi$, to each image after which white noise is added producing noisy measurements, $\bm{y}$, at a specified SNR. This produces a dataset $\mathcal{D}_{N_s} = \{(\overline{\bm{y}}_i, \overline{\bm{c}}_i): i = 1, 2, \ldots, N_s\}$ where $N_s$ is the number of data samples. Note that we randomly sample across all classes of images from CIFAR10 and CalTech101 in forming the training, validation, and testing datasets.

We consider two types of sensing matrices:
\begin{enumerate}
    \item Radon Transform: Typical in tomography, specifically X-ray computed tomography, we form sensing matrices, $\Psi$, corresponding to Radon transforms at a number of uniformly spaced angles.
    \item Random Gaussian Matrix: Typical in CS, $\Psi\in\mathbb{R}^{m\times n}$ has entries sampled from a standard Gaussian distribution with a given sampling ratio defined at $\frac{m}{n}$. We form three sensing matrices corresponding to sampling ratios of $0.5, 0.3,$ and $0.1$.
\end{enumerate}
Finally, we consider measurements with two different SNRs of 60dB or 40dB.

\subsection{Main Results} \label{sec:low training results}

%%%%%%%%%%%%%%%%%%%% 128x128 Radon, 60 angles %%%%%%%%%%%%%%%%%

\begin{figure*}
\centering
\begin{subfigure}{.19\linewidth}
    \centering
    \includegraphics[width=\linewidth]{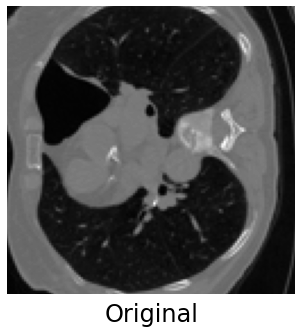}
\end{subfigure}
\begin{subfigure}{.19\linewidth}
    \centering
    \includegraphics[width=\linewidth]{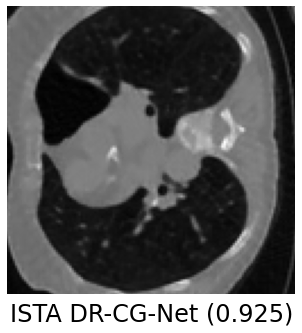}
\end{subfigure}
\begin{subfigure}{.19\linewidth}
    \centering
    \includegraphics[width=\linewidth]{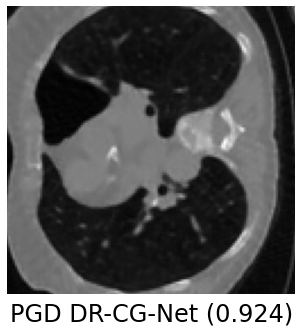}
\end{subfigure}
\begin{subfigure}{.19\linewidth}
    \centering
    \includegraphics[width=\linewidth]{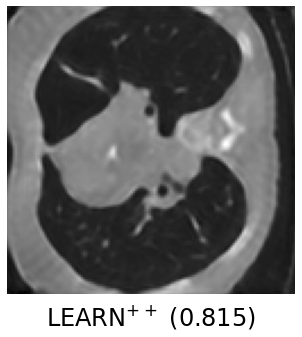}
\end{subfigure}
\begin{subfigure}{.19\linewidth}
    \centering
    \includegraphics[width=\linewidth]{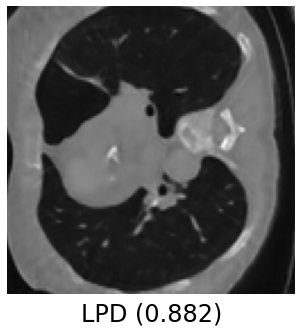}
\end{subfigure}

\begin{subfigure}{.19\textwidth}
    \centering
    \includegraphics[width=\linewidth]{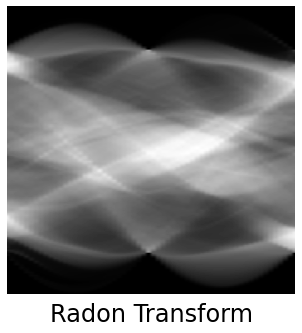}
\end{subfigure}
\begin{subfigure}{.19\linewidth}
    \centering
    \includegraphics[width=\linewidth]{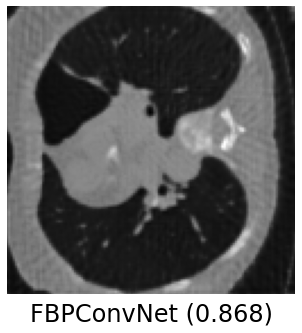}
\end{subfigure}
\begin{subfigure}{.19\linewidth}
    \centering
    \includegraphics[width=\linewidth]{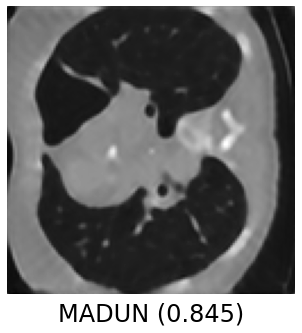}
\end{subfigure}
\begin{subfigure}{.19\linewidth}
    \centering
    \includegraphics[width=\linewidth]{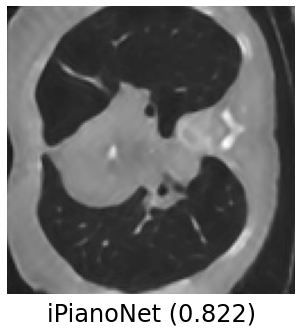}
\end{subfigure}
\begin{subfigure}{.19\linewidth}
    \centering
    \includegraphics[width=\linewidth]{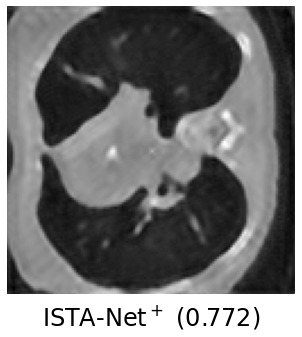}
\end{subfigure}
\caption{Image reconstructions (SSIM) using our DR-CG-Net and six competitive deep learning methods on a $128\times 128$ test scan after training with only 20 samples. The sensing matrix, $\Psi$, is a Radon transform at 60 uniformly spaced angles, $\Phi = I$, and each measurement has an SNR of 60dB. Our DR-CG-Net methods perform best visually and by SSIM.}
\label{fig:radon reconstructions128 60 angles}
\end{figure*}

%%%%%%%%%%%%%%%%%%%%%%%%%%%%%%%%%%%%%%%%%%%%%%

For each set of data discussed in Section~\ref{sec:data}, we train an ISTA DR-CG-Net, PGD DR-CG-Net, and each of the ten comparison methods, described at the beginning of this section, using varying amounts of training dataset sizes. For CIFAR10 images, the training datasets consist of $N_s  = 20, 100, 500, 1000,$ and $2000$ data samples and, after training, 8000 test data samples are provided to every network to assess its performance. From varying the amount of training data, we find that a core utility of DR-CG-Net is its superior performance on small training datasets, i.e. a highly underdetermined system in (\ref{eqn:loss}). As such, for CalTech101 and LoDoPaB-CT images, we use training datasets with $N_s = 20$ samples, to focus on evaluating low training scenarios in greater detail, and 200 test data samples are provided to every network after training.

Average PSNR and SSIM quality metrics, together with associated $99\%$ confidence intervals, on the test dataset reconstructions are used to evaluate network performance where higher values of these metrics correspond to reconstructed images that more closely match the original images. Note that the test data samples are formed from the same sensing matrix and noise level that produced the training data.

We remark that every method was trained using early stopping. That is, as shown in Fig.~\ref{supmat-fig:loss curves} in the supplementary material, training was conducted until the model initially overfits as compared to a validation dataset. In doing so, we ensure every model is sufficiently trained while also not being over-trained thereby presenting the best performance for each model for the provided set of training data.

For training datasets of size $20$ and $100$, we set $P_u$ as a scaled identity covariance matrix and a tridiagonal covariance matrix is used for all other training datasets. Next, for images larger than $64\times 64$ we approximate the Tikhonov solution with 100 NAGD steps as detailed in Fig.~\ref{fig:DR-CG-Net u GD block}. Lastly, for Radon inversion experiments, we take $\Phi = I$ allowing DR-CG-Net to intrinsically learn the optimal signal representation basis.

Shown in Fig.~\ref{fig:compare} is the average SSIM quality over a set of test CIFAR10 image reconstructions from 60dB or 40dB SNR Radon transform measurements when varying the amount of training data. As ReconNet and iRadonMAP perform significantly lower they are omitted from Fig.~\ref{fig:compare}. Additionally, as PGD DR-CG-Net performs equivalently to ISTA DR-CG-Net it too was omitted from Fig.~\ref{fig:compare}. Note that Fig.~\ref{supmat-fig:compare} in the supplementary material displays similar plots for PSNR quality. We can see from Fig.~\ref{fig:compare}, and Fig.~\ref{supmat-fig:compare} in the supplementary material, that DR-CG-Net outperforms, or performs comparably to, all compared methods in every training scenario and excels significantly in the lowest training scenarios of $N_s = 20$ and $N_s = 100$. However, the improvement of DR-CG-Net, over the comparison methods, diminishes when a greater amount of measurement noise is present and as the amount of training data increases. In particular, we observe from Fig.~\ref{fig:compare} that with enough training data and higher noise, DR-CG-Net can be matched by the best comparison method.

\begin{figure*}
\centering
\begin{subfigure}{.19\linewidth}
    \centering
    \includegraphics[width=\linewidth]{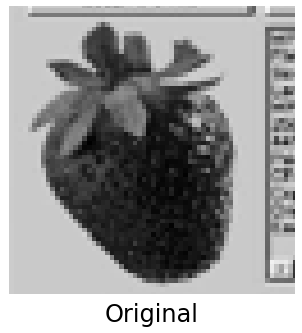}
\end{subfigure}
\begin{subfigure}{.19\linewidth}
    \centering
    \includegraphics[width=\linewidth]{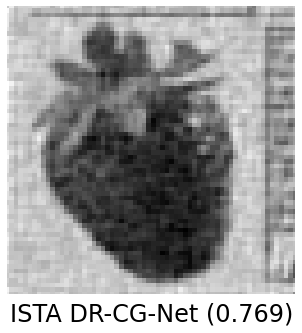}
\end{subfigure}
\begin{subfigure}{.19\linewidth}
    \centering
    \includegraphics[width=\linewidth]{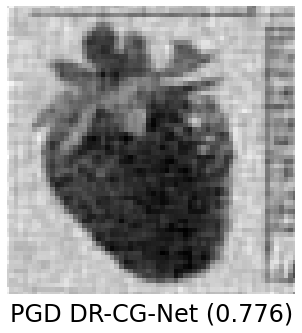}
\end{subfigure}
\begin{subfigure}{.19\linewidth}
    \centering
    \includegraphics[width=\linewidth]{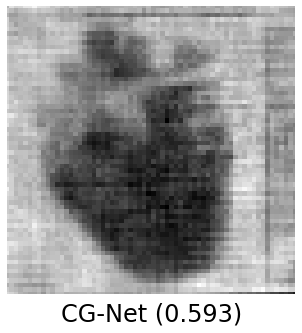}
\end{subfigure}
\begin{subfigure}{.19\linewidth}
    \centering
    \includegraphics[width=\linewidth]{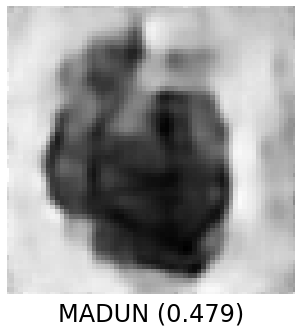}
\end{subfigure}

\begin{subfigure}{.19\textwidth}
    \centering
    \includegraphics[width=\linewidth]{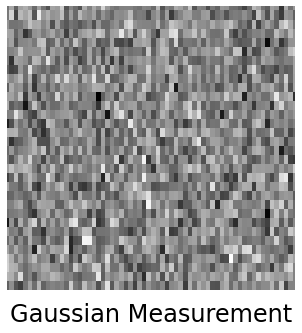}
\end{subfigure}
\begin{subfigure}{.19\linewidth}
    \centering
    \includegraphics[width=\linewidth]{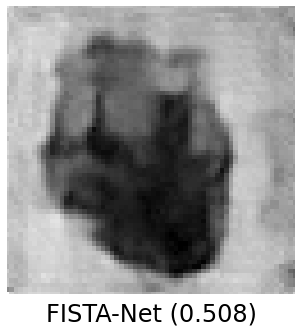}
\end{subfigure}
\begin{subfigure}{.19\linewidth}
    \centering
    \includegraphics[width=\linewidth]{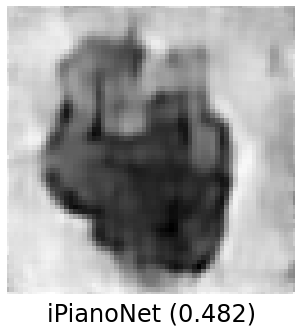}
\end{subfigure}
\begin{subfigure}{.19\linewidth}
    \centering
    \includegraphics[width=\linewidth]{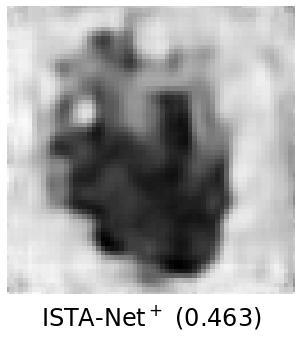}
\end{subfigure}
\begin{subfigure}{.19\linewidth}
    \centering
    \includegraphics[width=\linewidth]{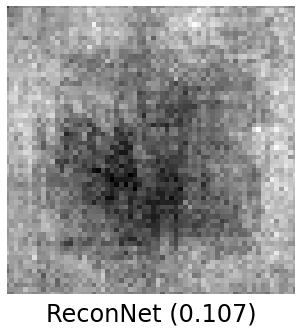}
\end{subfigure}
\caption{Image reconstructions (SSIM) using our DR-CG-Net and six comparative deep learning methods on a $64\times 64$ strawberry image after training on only 20 samples. $\Psi =$ Gaussian matrix at 0.5 sampling ratio, $\Phi = $ discrete cosine transformation, and measurements have an SNR of 60dB. DR-CG-Net method performs best visually, particularly the leaf detail, and by SSIM.}
\label{fig:0.5 ratio reconstructions}
\end{figure*}

\begin{figure*}
\centering
\begin{subfigure}{\textwidth}
\centering
\begin{subfigure}{.20\textwidth}
    \centering
    \includegraphics[width=\textwidth]{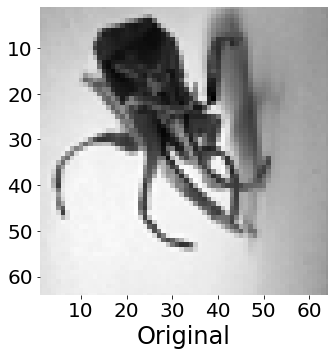}
\end{subfigure}
\begin{subfigure}{.20\textwidth}
    \includegraphics[width=\textwidth]{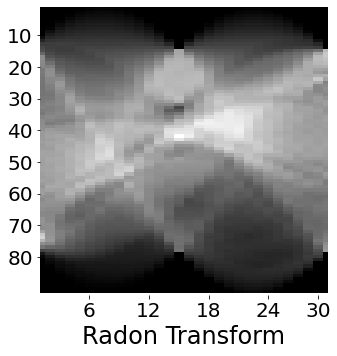}
\end{subfigure}
\begin{subfigure}{.20\textwidth}
    \centering
    \includegraphics[width=\textwidth]{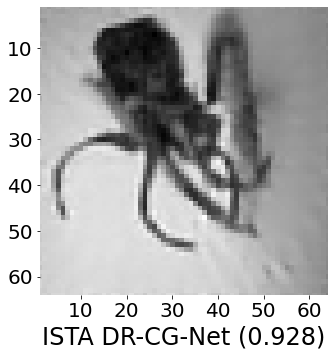}
\end{subfigure}
\begin{subfigure}{.20\textwidth}
    \centering
    \includegraphics[width=\textwidth]{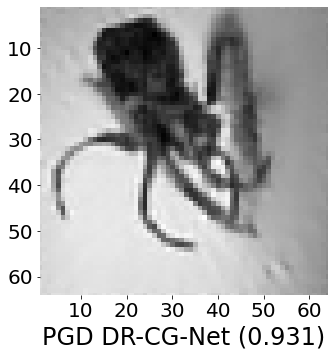}
\end{subfigure}
\end{subfigure}
\begin{subfigure}{.160\textwidth}
    \centering
    \includegraphics[width=\textwidth]{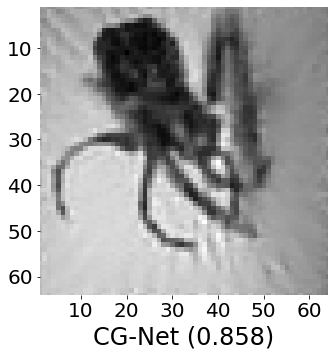}
\end{subfigure}
\begin{subfigure}{.160\textwidth}
    \centering
    \includegraphics[width=\textwidth]{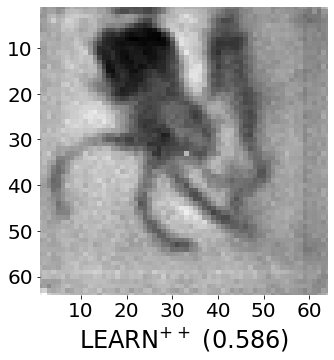}
\end{subfigure}
\begin{subfigure}{.160\textwidth}
    \centering
    \includegraphics[width=\textwidth]{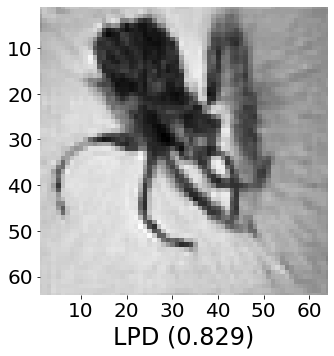}
\end{subfigure}
\begin{subfigure}{.160\textwidth}
    \centering
    \includegraphics[width=\textwidth]{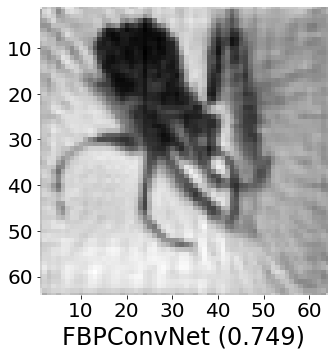}
\end{subfigure}
\begin{subfigure}{.160\textwidth}
    \centering
    \includegraphics[width=\textwidth]{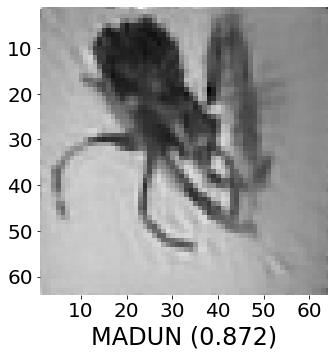}
\end{subfigure}
\begin{subfigure}{.160\textwidth}
    \centering
    \includegraphics[width=\textwidth]{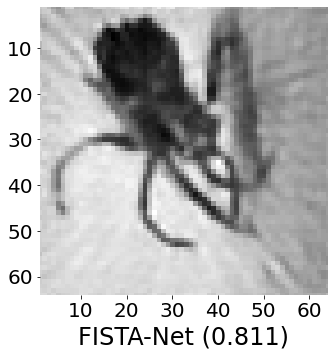}
\end{subfigure}
\begin{subfigure}{.160\textwidth}
    \centering
    \includegraphics[width=\textwidth]{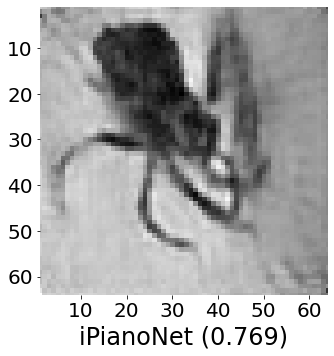}
\end{subfigure}
\begin{subfigure}{.160\textwidth}
    \centering
    \includegraphics[width=\textwidth]{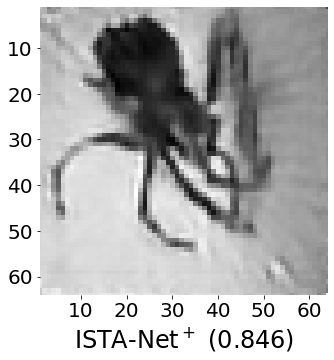}
\end{subfigure}
\begin{subfigure}{.160\textwidth}
    \centering
    \includegraphics[width=\textwidth]{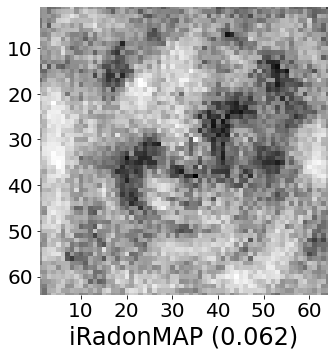}
\end{subfigure}
\begin{subfigure}{.160\textwidth}
    \centering
    \includegraphics[width=\textwidth]{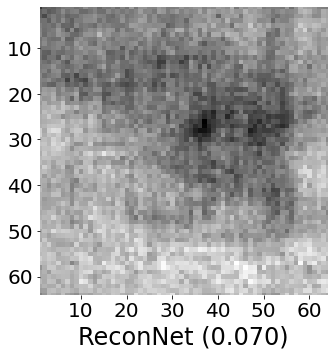}
\end{subfigure}
\begin{subfigure}{.160\textwidth}
    \centering
    \includegraphics[width=\textwidth]{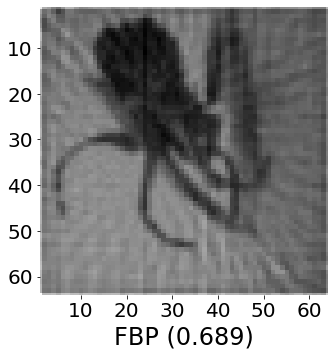}
\end{subfigure}
\begin{subfigure}{.160\textwidth}
    \centering
    \includegraphics[width=\textwidth]{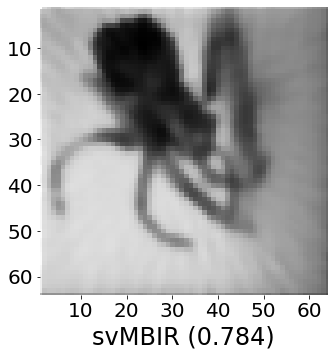}
\end{subfigure}
\caption{Image reconstructions (SSIM) using our DR-CG-Nets, ten comparative deep learning methods, and two baseline iterative-based methods on a $64\times 64$ spider image. The sensing matrix, $\Psi$, is a Radon transform at 30 uniformly spaced angles, $\Phi = I$, and each measurement has an SNR of 60dB. Our DR-CG-Net methods perform best visually and by SSIM.}
\label{fig:reconstructions64}
\end{figure*}

A visual comparison of all methods after training on only $N_s = 20$ samples is provided in Fig.~\ref{fig:radon reconstructions128 60 angles}, Fig.~\ref{fig:0.5 ratio reconstructions}, and Fig.~\ref{fig:reconstructions64}. In Fig.~\ref{fig:radon reconstructions128 60 angles} are the estimates of a $128\times 128$ scan from a Radon transform at 60 uniformly spaced angles with 60dB SNR. Fig.~\ref{fig:0.5 ratio reconstructions} displays the reconstructions of a $64\times 64$ test image from a Gaussian measurement at 0.5 sampling ratio. Lastly, Fig.~\ref{fig:reconstructions64} shows the estimates of a $64\times 64$ test image from a Radon transform at 30 uniformly spaced angles. Additional visualizations, after training on $N_s = 20$ LoDoPaB-CT and CIFAR10 samples, are presented in Fig.~\ref{supmat-fig:radon reconstructions128} and Fig.~\ref{supmat-fig:0.5 ratio reconstructions} in the supplementary material. Fig.~\ref{supmat-fig:radon reconstructions128} and Fig.~\ref{supmat-fig:0.5 ratio reconstructions} display reconstructed images from a 76 angle Radon transform and from a Gaussian measurement with 0.5 sampling ratio, respectively. In each case, we see that ISTA and PGD DR-CG-Net perform comparably with both producing superior reconstructions, both visually and by SSIM, to all compared methods.

\begin{figure*}[!t]
\centering
\begin{subfigure}{\textwidth}
    \centering
\begin{subfigure}{.32\linewidth}
    \centering
    \includegraphics[width=\linewidth]{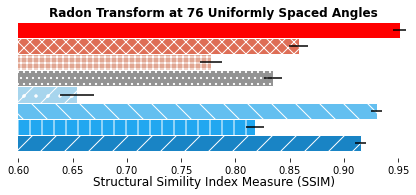}
\end{subfigure}
\begin{subfigure}{.32\linewidth}
    \centering
    \includegraphics[width=\linewidth]{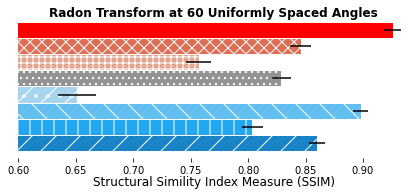}
\end{subfigure}
\begin{subfigure}{.32\linewidth}
    \centering 
    \includegraphics[width=\linewidth]{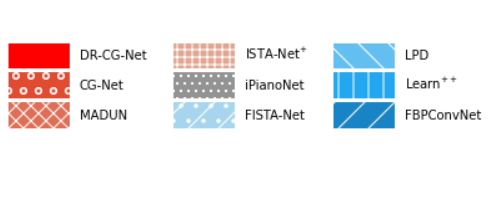}
\end{subfigure}
    \caption{Average SSIM, with $99\%$ confidence intervals, for nine deep learning-based image estimation methods reconstructing 200, $128\times 128$ LoDoPaB-CT images~\cite{lodopabct} from Radon transform measurements. Each measurement has an SNR of 60dB and $\Phi = I$.}
    \label{fig:low training radon128}
\end{subfigure}

\begin{subfigure}{\textwidth}
    \centering
\begin{subfigure}{.32\linewidth}
    \centering
    \includegraphics[width=\linewidth]{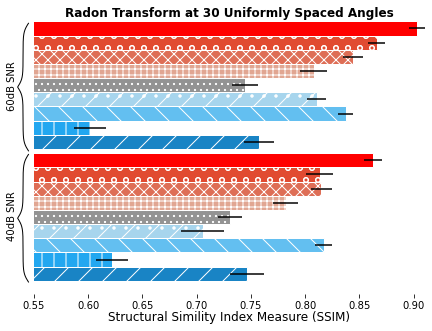}
\end{subfigure}
\begin{subfigure}{.32\linewidth}
    \centering
    \includegraphics[width=\linewidth]{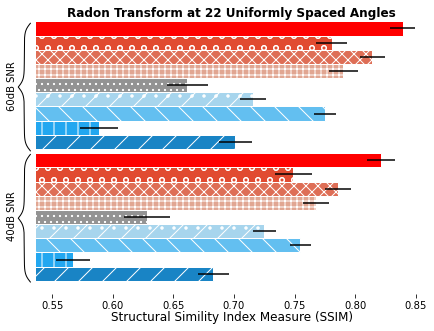}
\end{subfigure}
\begin{subfigure}{.32\linewidth}
    \centering 
    \includegraphics[width=\linewidth]{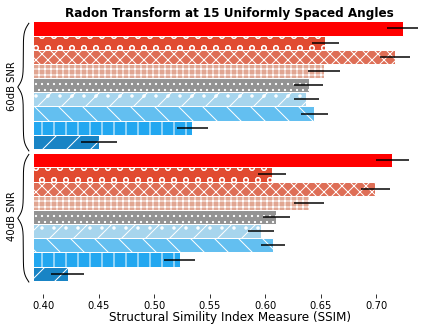}
\end{subfigure}
    \caption{Average SSIM, with $99\%$ confidence intervals, for nine deep learning-based image estimation methods reconstructing 200, $64\times 64$ CalTech101 images~\cite{CalTech101} from Radon transform measurements. Each measurement has an SNR of 60dB or 40dB and $\Phi = I$.}
    \label{fig:low training radon64}
\end{subfigure}

\begin{subfigure}{\textwidth}
    \centering
\begin{subfigure}{.32\linewidth}
    \centering
    \includegraphics[width=\linewidth]{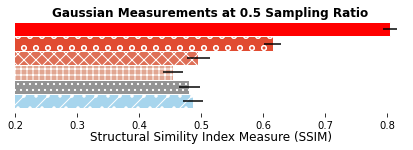}
\end{subfigure}
\begin{subfigure}{.32\linewidth}
    \centering
    \includegraphics[width=\linewidth]{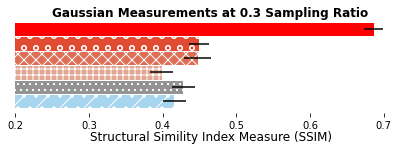}
\end{subfigure}
\begin{subfigure}{.32\linewidth}
    \centering
    \includegraphics[width=\linewidth]{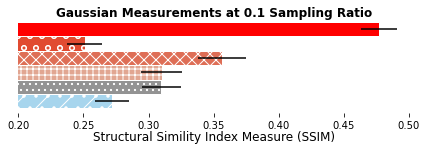}
\end{subfigure}
    \caption{Average SSIM, with $99\%$ confidence intervals, for six deep learning-based image estimation methods reconstructing 200, $64\times 64$ CalTech101 images from Gaussian measurements with an SNR of 60dB and $\Phi = $ discrete cosine transformation.}
    \label{fig:low training CS}
\end{subfigure}
\caption{Test image reconstruction quality for deep learning-based image estimation methods trained on only 20 samples. \textbf{In all cases, our method, DR-CG-Net given by the top solid red bar, outperforms the other approaches.}}
\label{fig:low training}
\end{figure*}

Fig.~\ref{fig:low training} further highlights the excellent performance of DR-CG-Net over the competitive methods in low training scenarios by providing the average SSIM over 200 test images, with $99\%$ confidence intervals, after training only with $N_s = 20$ LoDoPaB-CT or CalTech101 images. In particular, Fig.~\ref{fig:low training radon128} and Fig.~\ref{fig:low training radon64} show the SSIM performance of all methods for reconstructing test images from Radon transform measurements. Fig.~\ref{fig:low training CS} shows the SSIM performance of all methods for reconstructing test images from Gaussian measurements. Observing Fig.~\ref{fig:low training}, DR-CG-Net outperforms each prior art method, in both Radon inversion and CS tasks when training on limited datasets. We remark that PGD DR-CG-Net performs similarly to ISTA DR-CG-Net and is thus omitted from Fig.~\ref{fig:low training}. As ReconNet and iRadonMAP perform the lowest, they are omitted from Fig.~\ref{fig:low training}. Additionally, CG-Net is omitted from Fig.~\ref{fig:low training radon128} due to an insurmountable computational cost to train on $128\times 128$ images. Lastly, as LPD, LEARN$^{++}$, and FBPConvNet are CT-specific methods, the CS problem is not applicable and so these comparisons are omitted from Fig.~\ref{fig:low training CS}.

An advantage of DR-CG-Net, as highlighted by our CIFAR10 and CalTech101 results in Fig.~\ref{fig:compare}, Fig.~\ref{fig:low training}, Fig.~\ref{fig:0.5 ratio reconstructions}, and Fig.~\ref{fig:reconstructions64} is its ability to perform well on non-uniform data. Both CIFAR10 and CalTech101 contain many distinct classes of images from which we uniformly sample to create our training and testing datasets. In training with only 20 samples, few if any samples from each class will be seen by the network. Despite this, DR-CG-Net is able to recover high quality images in the test dataset that contains many more samples from both the classes present in the training data and from classes not present in the training data. That is, DR-CG-Net has the ability to learn only from a handful of samples in a class and generalize well to unseen classes of samples.

We posit that the high performance of DR-CG-Net, especially in low training scenarios, is due to the natural incorporation of the powerful CG prior through unrolling G-CG-LS. Specifically, the Tikhonov estimation layers and Hadamard product layer provide significant data consistency structure by closely matching input measurements and measured estimated signals. Through this data consistency, a natural regularization for the DNN optimization is enforced by restricting the possible generated signals to a CG class. A thorough examination of the interplay between the iterative algorithm regularization and regularization for the DNN optimization is an intriguing subject of future work for unrolled DNN methods as a whole. Finally, the ability to learn the scale variable distribution, unlike in CG-Net where it is fixed as log-normal, provides DR-CG-Net with greater training capacity and flexibility to edge out CG-Net in test reconstruction performance in low training and significantly outperform CG-Net in higher training.

Lastly, we reiterate that the results of Fig.~\ref{fig:radon reconstructions128 60 angles},~\ref{fig:low training},~\ref{fig:0.5 ratio reconstructions},~\ref{fig:reconstructions64}, and in the supplementary material are for training on limited datasets, which is of interest in many applications in tomographic imaging. Consequently, the results presented are not representative of the performance of these methods with unlimited training data where our method is likely to be matched by the comparison methods as indicated by the results in Fig.~\ref{fig:compare}.

\subsection{Computational Time  and Complexity}

Table~\ref{tab:computational time} lists the average computational time per image, in milliseconds, across 8000 test image reconstructions running on a 64-bit Intel(R) Xeon(R) CPU E5-2690. Among the competing methods, we see that DR-CG-Net is comparable in reconstruction time to MADUN. While CG-Net provided the second-best results in low training scenarios, we can see from Table~\ref{tab:computational time} that DR-CG-Net is, by comparison, at least 20 times faster on average in reconstructing images.

 The speed increase of DR-CG-Net over CG-Net is from three main sources: First, each update of $\bm{z}$ in CG-Net implements an eigendecomposition calculation, which is slower than each $\bm{z}$ update in DR-CG-Net requiring only matrix-vector products. Second, CG-Net was formed by unrolling a larger number of iterations, specifically $(K, J) = (20, 1)$, whereas DR-CG-Net only requires $(K, J) = (3, 4)$ for excellent performance. Third, the Woodbury matrix identity is employed in DR-CG-Net to accelerate calculating the Tikhonov solution.

\begin{table}[!b]
\caption{Average reconstruction time of $32\times 32$ images from Radon transform measurements at 15 uniform angles.}
    \centering
    \begin{tabular}{c|c|c|c|c}
    Method     &  PGD DR-CG-Net & ISTA DR-CG-Net & CG-Net & MADUN  \\
    Time (ms)     & 22.0 & 52.7 & 765 & 26.0 \\
    \hline
        Method  & ReconNet & iRadonMAP & LEARN$^{++}$  & LPD   \\
    Time (ms)  & $0.65$ & 4.5 & 10.6 & 4.7 \\
    \hline
       Method  & FBPConvNet & iPiano-Net & FISTA-Net & ISTA-Net$^+$ \\
    Time (ms)  & 2.0 & 13.0 & 6.4 & 7.9
    \end{tabular}
    \label{tab:computational time}
\end{table}

To explore the scalability of DR-CG-Net we consider the computational complexity of a forward pass through the network, in floating-point operations (FLOPs), with respect to the signal of interest size, $n$, and measurement size, $m$. We assume a batch size of one as the complexity simply grows proportionally to batch size. Block $\mathcal{Z}_0$ requires one matrix-vector product using $2mn$ FLOPs. For $i\geq 1$, block $\mathcal{Z}_i$ requires one data fidelity gradient update, $r_k^{(j)}$, via (\ref{eqn:data fidelity gradient}) using $4mn+4n+m$ FLOPs, $D$ convolutions with $3\times3$ kernels and $f_1, \ldots, f_D$ filter channels using $18n\sum_{d=1}^D f_d$ FLOPs, and finally one addition of length $n$ vectors using $n$ FLOPs.

In the exact calculation of the Tikhonov solution via (\ref{eqn:Tikhonov solution rewritten}), first $I+A_{\bm{z}}P_u A_{\bm{z}}^T$ is calculated using $2mn^2 + 2m^2n + mn + m$ FLOPs. Second, a $m\times m$ system of equations is solved, naively using $\frac{2}{3}m^3 +\mathcal{O}(m^2)$ FLOPs. Finally, a $m\times n$ matrix vector product and $n\times n$ matrix vector product are performed using $2mn$ and $2n^2$ FLOPs, respectively. Thus, a total of $\frac{2}{3}m^3+2mn^2+\mathcal{O}(m^2n+n^2)$ FLOPs are used in calculating (\ref{eqn:Tikhonov solution rewritten}). Therefore, when we calculate the Tikhonov solution exactly, the total FLOPs for DR-CG-Net scales as
\begin{align*}
    \mathcal{O}\left(KJ\left[m+\scalebox{1}{$\sum_{d=1}^D$} f_d\right]n + Km^3 + Kmn^2\right).
\end{align*}

When we consider, instead, the approximation of the Tikhonov solution by $J_u$ NAGD steps, the first two evaluations of $r_u$ via (\ref{eqn:G-CG-LS GD u step}) are computed, using $2n^2 + 4mn + 5n+m$ FLOPs each. Second, a length-$n$ vector subtraction, scalar multiplication, and length-$n$ vector addition are calculated using $3n$ FLOPs. Therefore, when we approximate the Tikhonov solution with NAGD steps, the total FLOPs for DR-CG-Net scales as
\begin{align*}
    \mathcal{O}\left(KJ\left[m+\scalebox{1}{$\sum_{d=1}^D$} f_d\right]n + KJ_un^2\right).
\end{align*}

Thus, computational complexity is reduced by a power of $m$ for the replacement of the exact Tikhonov solution with NAGD steps. Still, both versions have non-linear polynomial growth in computational complexity and thus, further computational reductions are required for scaling DR-CG-Net to signals of significantly higher dimension.

%%%%%%%%%%%%%%%%%%%% 64x64 Gaussian %%%%%%%%%%%%%%%%%

%%%%%%%%%%%%%%%%%%%%%%%%%%%%%%%%%%%%%%%%%%%%%%

\subsection{Network Parameters}

\begin{table}[!b]
\caption{Parameter count for each compared deep learning method when reconstructing a $32\times 32$ image from Radon transform measurements at 15 uniformly spaced angles.}
    \centering
    \adjustbox{max width=\columnwidth}{
    \begin{tabular}{c|c||c|c}
        Method  & Parameters ($\times 10^5$) & Method & Parameters ($\times 10^5$) \\
        \hline\hline
        ISTA DR-CG-Net & 7.26 & PGD DR-CG-Net & 7.26 \\
        CG-Net & 1.17 & ISTA-Net$^+$ & 3.37 \\
        LPD  & 2.53 & MADUN & 29.7\\
        LEARN$^{++}$ & 12.0 & FISTA-Net & 0.75\\
        iRadonMAP & 8.33 & iPiano-Net & 19.3\\
        FBPConvNet & 7.09 & ReconNet & 7.30
    \end{tabular}
    }
    \label{table:parameters}
\end{table}

Table \ref{table:parameters} displays the number of parameters, i.e. unknowns, for DR-CG-Net and all ten compared deep learning-based methods for linear inverse problems. Note that the ratio of unknowns to data point measurements can be obtained by dividing each entry of Table \ref{table:parameters} by the number of data points. For example, when training with 20 CIFAR10 images and 15 angle Radon transforms we use $20\times (\textnormal{image size} + \textnormal{Radon transform size}) = 20\times(1024+690) = 3.428\times 10^4$ points of data, giving a ratio of unknowns to measurements of $7.26\times 10^5/(3.428\times 10^4)\approx 21.2$ for DR-CG-Net. Instead, when training with 20 LoDoPaB-CT images and 60 angle Radon transforms, we have a ratio of unknowns to measurements of $7.26\times 10^5/(5.46\times 10^5) \approx 1.33$ for DR-CG-Net. These ratios are on par with FBPConvNet and ReconNet that have a similar number of parameters to DR-CG-Net.

While the ratio of unknowns to measurements is high in low-training scenarios, our empirical results demonstrate that DR-CG-Net handles this situation well. In particular, we note that despite DR-CG-Net having a significantly greater ratio of unknowns to measurements than CG-Net, DR-CG-Net outperforms CG-Net when both are trained on small training datasets.

\subsection{Refinement Block Study} \label{sec:refinement block}

Here, we investigate the impact that the refinement block, $\mathcal{G}$, has on the quality of the reconstructed signals produced by DR-CG-Net. To this end, we empirically evaluate two alternative formulations for DR-CG-Net where $\mathcal{G}$ is removed before training (b.t.) and after training (a.t.) as shown in Table~\ref{table:refinement block study}. When $\mathcal{G}$ is removed before training, a fresh DR-CG-Net is trained, as in the setup of Section~\ref{sec:low training results}, with the $C$ block now as output. Instead, when $\mathcal{G}$ is removed after training, a trained DR-CG-Net from Section~\ref{sec:low training results} is truncated to return the $C$ block as output and no additional training is conducted. 

For both instances of removing the refinement block, the reconstruction quality of DR-CG-Net remains significant indicating that the unrolled G-CG-LS portion of DR-CG-Net is the core component in the superb reconstructions produced by DR-CG-Net. Additionally, higher noise, i.e. 40dB SNR, results in a larger discrepancy in the DR-CG-Net performance when the refinement block is or is not implemented as compared to the lower noise case. This suggests that the refinement block primarily denoises the estimated image from the unrolled G-CG-LS portion of DR-CG-Net.

\begin{table}[!ht]
    \caption{Refinement block study for ISTA DR-CG-Net. Displayed is the average SSIM and PSNR for CIFAR10 image reconstructions from a Radon transform, at several different amounts of uniformly spaced angles, with a set SNR. \textbf{T} and \textbf{F} indicate if the refinement block is or is not used, respectively. Further, b.t. and a.t. denote if the refinement block is removed before or after training, respectively.}
    \label{table:refinement block study}
    \centering
    \begin{tabular}{c|c|c|c|c|c|c|c}
    && \multicolumn{6}{c}{Training Dataset Size} \\
     $\mathcal{G}$    & (Angles,SNR) & \multicolumn{2}{c|}{20} & \multicolumn{2}{c|}{100} & \multicolumn{2}{c}{500} \\
     && SSIM & PSNR & SSIM & PSNR & SSIM & PSNR \\
     \hline
      \textbf{F} b.t.   & & 0.901 & 28.22 & 0.922 & 28.96 & 0.937 & 29.89 \\
      \textbf{F} a.t. & (15,60) & 0.900 & 27.97 & 0.912 & 28.35 & 0.917 & 28.76 \\
     \textbf{T} & & 0.919 & 28.59 & 0.932 & 29.45 & 0.940 & 30.07 \\
           \hline
    \textbf{F} b.t.   & & 0.852 & 25.98 & 0.882 & 26.94 & 0.897 & 27.71 \\
      \textbf{F} a.t. & (15,40) & 0.857 & 26.10 & 0.862 & 26.24 & 0.869 & 26.30 \\
     \textbf{T} & & 0.891 & 26.86 & 0.903 & 27.47 & 0.909 & 27.87\\
           \hline
    \textbf{F} b.t.   & & 0.821 & 25.13 & 0.840 & 25.60 & 0.855 & 26.12\\
      \textbf{F} a.t. & (10,60) & 0.820 & 25.03 & 0.829 & 25.30 & 0.839 & 25.63 \\
     \textbf{T} & & 0.828 & 25.17 & 0.844 & 25.69 & 0.860 & 26.29 \\
                 \hline
    \textbf{F} b.t.   & &  0.775 & 23.84 & 0.794 & 24.34 & 0.826 & 25.15 \\
      \textbf{F} a.t. & (10,40) & 0.775 & 23.87 & 0.787 & 24.12 & 0.789 & 24.27\\
     \textbf{T} & & 0.811 & 24.49 & 0.827 & 25.00 & 0.839 & 25.44 \\
    \end{tabular}
\end{table}

%%%%%%%%%%%%%%%%%%%%% 64x64 DNN Recons %%%%%%%%%%%%%%%%%
%%%%%%%%%%%%%%%%%%%%%%%%%%%%%%%%%%%%%%%%%%%%%%%%%%%%%%%%%%%%%%%%%%%%
 
\section{Conclusion and Future Work}

Using a powerful learned CG class of densities, that subsume many commonly used priors in imaging and CS, this paper has presented new fundamental techniques for linear inverse problems. We developed a novel CG-based iterative estimation algorithm, named G-CG-LS, that allows for problem-specific choices of the scale variable distribution by generalizing on prior CG-based methods. Applying algorithm unrolling to G-CG-LS, we constructed a novel CG-based deep neural network, named DR-CG-Net, that optimally learns the scale variable portion and the Gaussian covariance portion of the CG distribution. Hence, DR-CG-Net has the flexibility to learn the prior while constraining to the informative CG class of distributions.

We conducted a theoretical characterization of G-CG-LS and a fundamental numerical validation of DR-CG-Net in tomographic imaging and CS problems. Across multiple datasets, we empirically demonstrated that DR-CG-Net significantly outperforms competitive state-of-the-art deep learning-based methods in tomographic imaging and CS scenarios, especially in the difficult case of low-training. While CG-Net, which DR-CG-Net expands upon, is the closest comparative method in signal reconstruction quality for the low-training scenarios, CG-Net is still appreciably outperformed by DR-CG-Net both in estimated signal quality and in the computational time required to train and reconstruct a signal.

Leveraging the new foundations established in this paper, numerous opportunities for future investigation are illuminated including: analyzing the performance of DR-CG-Net on larger signal reconstructions, on data corrupted by different noise models, and for training with larger datasets. Studying each of these future objectives will provide greater insight into the practical, real-world applicability of our method.

We foresee the main complication in reconstructing larger images to be hardware constraints in the memory required to store matrices for and calculate the Tikhonov solution. One workaround is replacing the Tikhonov solution with gradient descent steps, as we did for our $128\times 128$ image results. While gradient descent steps only approximate the Tikhonov solution, we empirically showed that DR-CG-Net still successfully estimates images and outperforms all compared state-of-the-art approaches. For Radon transform measurements, $A$, in (\ref{eqn:linear_msrmt}), is a sparse matrix that can be stored in a memory-efficient way (i.e. as a SparseTensor object in TensorFlow). Thus, with the development of a sparse linear solver through TensorFlow, only a slight modification of our code is necessary to efficiently deploy DR-CG-Net, with the exact Tikhonov solution, on larger images. Additionally, for Radon transform measurements, a matrix-free method, such as conjugate gradient, may be substituted to approximate the Tikhonov solution without constructing and storing the forward operator matrix. For CS applications, a common technique is to split larger images into disjoint blocks of small size and then measure and reconstruct the blocks separately~\cite{MADUN, song2023MAPUN}; this strategy can be employed for future experimental evaluation of DR-CG-Net.

Another intriguing line of work is empirically studying DR-CG-Net, and each comparison deep learning method, for training and testing on a single category of images. The training and testing datasets employed in this paper contain images from multiple categories, e.g. CIFAR10 has 10 categories and CalTech101 has 101 categories, creating a significant statistical variability in the datasets. For some methods, such as FBPConvNet, the statistical variability in the training and testing datasets can distort performance. As such, it would be of interest to observe if DR-CG-Net continues to provide state-of-the-art performance on single modal datasets or if other deep learning methods are better suited when homogeneity in the training and testing datasets is known upfront. Furthermore, it would be of interest to observe how DR-CG-Net and all the other deep learning methods generalize to images from alternative categories after training on a single category. We conjecture that DR-CG-Net will generalize well to images outside of the training class in a similar manner to how well it performs on the CalTech101 dataset.

Additional future work can explore alternative scale variable update methods, rather than PGD and ISTA, such as the alternating direction method of multipliers. The unrolled implementation for DR-CG-Net would require careful construction when using another scale variable update method but may provide performance benefits in reconstructed image quality. Furthermore, convergence properties of G-CG-LS with alternative scale variables update methods could be proven.

Finally, on the theoretical front, an open question emerging from our work is the formulation of a mathematical framework to explain the excellent generalization properties of DR-CG-Net in low-training scenarios. Approaches from statistical learning theory literature, such as constructing generalization error bounds, may provide insight into the superb low-training performance of DR-CG-Net.

\section{Appendix} \label{apndx:proofs}

First, the equivalence of a MAP estimate and our G-CG-LS estimate, which is proved in the supplementary material.
\begin{proposition} \label{prop:MAP_estimate}
    The optimum of cost function $F$ in (\ref{eqn:iterative_cost_func}) is equivalent to a MAP estimate of $\bm{z}$ and $\bm{u}$ from (\ref{eqn:linear_msrmt}).
\end{proposition}
Second, an optimality condition defining a stationary point.
\begin{deff} \label{def:optimality conditions}
Let $f:\mathcal{X}\to\mathbb{R}$ for convex $\mathcal{X}$. A point $\bm{x}^*\in\mathcal{X}$ is \textbf{stationary} if and only if for all $\bm{x}\in\mathcal{X}$ it holds that either
\begin{enumerate}
    \item when $f$ is differentiable, then $\langle \nabla f(\bm{x}^*), \bm{x}-\bm{x}^*\rangle \geq 0$
\end{enumerate}
or
\begin{enumerate}  
    \item[2)] when $f$ is convex, then there exists a $\bm{d}^*$ in the subdifferential $\partial f(\bm{x}^*)$ such that $\langle \bm{d}(\bm{x}^*), \bm{x}-\bm{x}^*\rangle \geq 0$.
\end{enumerate}
\end{deff}

\subsection{Projected Gradient Descent Bounds} \label{apndx:PGD bounds}

Recall for a differentiable function $f:\mathcal{X}\to\mathbb{R}$ and initial point $\bm{x}_0\in\mathcal{X}$ the PGD sequence $\{\bm{x}_k\}_{k = 0}^\infty$ is defined as
\begin{align*}
    \bm{x}_{k+1} = \mathcal{P}_{\mathcal{X}}(\bm{x}_{k} - \eta_{k+1}\nabla f(\bm{x}_{k})),
\end{align*}
where $\eta_{k+1}>0$ a step size and $\mathcal{P}_{\mathcal{X}}$ a projection onto $\mathcal{X}$~\cite{bertsekas2016nonlinearprogramming}. A backtracking linesearch in PGD chooses the largest $\eta_{k+1} \leq 1$ such that, for $\alpha \in (0, 1/2]$,
\begin{align}
f(\bm{x}_{k+1}) \leq f(\bm{x}_{k}) - \alpha \langle \nabla f(\bm{x}_k), \bm{x}_k - \bm{x}_{k+1}\rangle. \label{eqn:backtracking line search}
\end{align}

Next, we bound the cost function change for a PGD step.

\begin{lemma}\label{lemma:PGD step bound}
Let $f:\mathcal{X}\to \mathbb{R}$, for convex $\mathcal{X}\subseteq\mathbb{R}^d$, be a $C^2$ function with $L$-Lipschitz continuous gradient on a compact subset $\mathcal{S}\subseteq \mathcal{X}$. For $\bm{x}_0\in \mathcal{S}$ and some positive constant $c$, the PGD sequence $\{\bm{x}_k\}_{k = 0}^{\infty}$ with fixed step size $0 < \eta < 2/L $ or backtracking linesearch step size satisfies
\begin{align}
    f(\bm{x}_{k}) - f(\bm{x}_{k+1})\geq c \norm{\bm{x}_{k+1}-\bm{x}_{k}}_2^2. \label{eqn:projected gradient descent bound}
\end{align}
\end{lemma}
\begin{proof}
    From \cite{cheney1959proximity} $\mathcal{P}_{\mathcal{X}}(\bm{x})$ projects $\bm{x}$ onto $\mathcal{X}$ if and only if 
    \begin{align}
        \langle \bm{x}-\mathcal{P}_{\mathcal{X}}(\bm{x}), \bm{v}-\mathcal{P}_{\mathcal{X}}(\bm{x}) \rangle \leq 0 \hspace{.5cm} \textnormal{ for all } \bm{v}\in\mathcal{X}. \label{eqn:projection operator requirement}
    \end{align}
    In (\ref{eqn:projection operator requirement}) take $\bm{x} = \bm{x}_k - \eta_{k+1}\nabla f(\bm{x}_k)$ and $\bm{v} = \bm{x}_k$, since $\mathcal{P}_{\mathcal{X}}(\bm{x}_k - \eta_{k+1}\nabla f(\bm{x}_k)) \equiv \bm{x}_{k+1}$ we have
    \begin{align*}
      0 &\geq  \langle\bm{x}_k - \eta_{k+1}\nabla f(\bm{x}_k) - \bm{x}_{k+1}, \bm{x}_k - \bm{x}_{k+1}\rangle \\
        &= \norm{\Delta\bm{x}_k}_2^2 - \eta_{k+1}\langle\nabla f(\bm{x}_k), \Delta\bm{x}_k\rangle
    \end{align*}
    for $\Delta\bm{x}_k \coloneqq \bm{x}_k - \bm{x}_{k+1}$. Implying
    \begin{align}
        -\langle\nabla f(\bm{x}_k), \Delta\bm{x}_k\rangle &\leq -\norm{\Delta\bm{x}_k}_2^2/\eta_{k+1}. \label{eqn:projected gradient step bound}
    \end{align}
    
    As $\nabla f$ is $L$-Lipschitz we can bound the spectral radius of the Hessian of $f$ by $L$. Combining this with Taylor's theorem to a second order remainder produces
    \begin{align}
        f(\bm{x}_{k+1})&\leq f(\bm{x}_{k}) -\langle\nabla f(\bm{x}_{k}),\Delta \bm{x}_k\rangle + L\norm{\Delta\bm{x}_k}_2^2/2 \label{eqn:Lipschitz continuous gradient result}\\
        &\leq f(\bm{x}_{k}) -\norm{\Delta\bm{x}_k}_2^2/\eta_{k+1} +L\norm{\Delta\bm{x}_k}_2^2/2 \nonumber \\
        &= f(\bm{x}_{k})-\left(1/\eta_{k+1}-L/2\right)\norm{\Delta\bm{x}_k}_2^2 \label{eqn:Lipschitz continuous gradient bound} 
    \end{align}
    where the second inequality results from using (\ref{eqn:projected gradient step bound}). For fixed step sizes $\eta_{k+1} = \eta$ where $0 < \eta < 2/L$ we have (\ref{eqn:Lipschitz continuous gradient bound}) produces (\ref{eqn:projected gradient descent bound}) with $c = \frac{1}{\eta}-\frac{L}{2}.$  
    
    Next, combining (\ref{eqn:projected gradient step bound}) and (\ref{eqn:Lipschitz continuous gradient result})  again
    \begin{align*}
        f(\bm{x}_{k+1}) \leq f(\bm{x}_k) - \left(1-L\eta_{k+1}/2\right) \left\langle \nabla f(\bm{x}_k), \Delta\bm{x}_k\right\rangle,
    \end{align*}
    which implies the backtracking linesearch requirement (\ref{eqn:backtracking line search}) will hold when $\eta_{k+1}\leq 1/L$. Combining (\ref{eqn:backtracking line search}) and (\ref{eqn:projected gradient step bound}) with the fact that $\eta_{k+1}\leq 1$ produces (\ref{eqn:projected gradient descent bound}) with $c = \alpha$.
\end{proof}

Finally, we show that fixed points of PGD are stationary.
\begin{lemma} \label{lemma:PGD fixed points are stationary}
    PGD fixed points are stationary.
\end{lemma}
\begin{proof}
    Let $\bm{x}^*$ be a fixed point, i.e. $\bm{x}^* = \mathcal{P}(\bm{x}^*-\eta\nabla f(\bm{x}^*))$ for $\eta > 0$. Using (\ref{eqn:projection operator requirement}) with $\bm{x} = \bm{x}^*-\eta\nabla f(\bm{x}^*)$ shows that $\bm{x}^*$ satisfies $1)$ of Definition~\ref{def:optimality conditions} and is a stationary point.
\end{proof}

\subsection{Iterative Shrinkage and Thresholding Algorithm Bounds} \label{apndx:ISTA bounds}

Recall \cite{fast_ISTA} for a convex and differentiable function $f:\mathcal{X}\to\mathbb{R}$, a convex (possibly non-smooth) function $h:\mathcal{X}\to\mathbb{R}$, and $\bm{x}_0\in\mathcal{X}$, the ISTA sequence $\{\bm{x}_k\}_{k = 0}^\infty$ on $f(\bm{x})+h(\bm{x})$ is
\begin{align}
    \bm{x}_{k+1} &= \textnormal{prox}_{\eta_{k+1}h}(\bm{x}_{k} - \eta_{k+1}\nabla f(\bm{x}_{k})) \label{eqn:ISTA update step}
\end{align}
where $\eta_{k+1} > 0$ is a step size. As $h$ can be non-smooth, a backtracking linesearch in ISTA only requires the gradient of $f$ by choosing the largest $\eta_{k+1}\leq 1$ such that
\begin{align}
    \resizebox{\columnwidth}{!}{${\displaystyle f(\bm{x}_{k+1}) \leq f(\bm{x}_k) - \langle\nabla f(\bm{x}_k),\bm{x}_k-\bm{x}_{k+1}\rangle +\frac{\norm{\bm{x}_{k+1}-\bm{x}_{k}}_2^2}{2\eta_k}. }$} \label{eqn:backtracking linesearch ISTA}
\end{align}

Now we bound the cost function change for an ISTA step.

\begin{lemma}\label{lemma:ista step bound}
    Let $f:\mathcal{X}\to\mathbb{R}$ be convex and have $L$-Lipschitz continuous gradient on a compact subset $\mathcal{S}\subseteq \mathcal{X}$. Let $h:\mathcal{X}\to\mathbb{R}$ be convex. Then the ISTA sequence $\{\bm{x}_k\}_{k = 1}^\infty$ on $r \coloneqq f+h$ with fixed step size $\eta$, for $0<\eta\leq 1/L$, or step size determined by a backtracking linesearch, satisfies
    \begin{align}
    r(\bm{x}_{k}) - r(\bm{x}_{k+1})\geq c \norm{\bm{x}_{k+1}-\bm{x}_{k}}_2^2 \label{eqn:ista bound}
\end{align}
for some positive constant $c$.
\end{lemma}
\begin{proof}
     Let $\bm{p}_{\eta}(\bm{v}) = \textnormal{prox}_{\eta h}(\bm{v}-\eta\nabla f(\bm{v}))$ and define $\Delta \bm{v} \coloneqq \bm{p}_{\eta}(\bm{v}) - \bm{v}$. Since $f$ and $h$ are convex, from \cite{fast_ISTA} (Lemma 2.3) for any $\eta$ such that
    \begin{align}
        f(\bm{p}_{\eta}(\bm{v})) &\leq f(\bm{v}) + \langle\nabla f(\bm{v}), \Delta \bm{v}\rangle + \norm{\Delta\bm{v}}_2^2/(2\eta) \label{eqn:ista step req}
    \end{align}
    it holds that
    \begin{align}
        r(\bm{x}) - r(\bm{p}_{\eta}(\bm{v})) &\geq \norm{\Delta\bm{v}}_2^2/(2\eta) + \langle \bm{v}-\bm{x}, \Delta\bm{v} \rangle/\eta. \label{eqn:ista step consequence}
    \end{align}
    As $f$ has $L$-Lipschitz continuous gradient, then (\ref{eqn:ista step req}), and thus (\ref{eqn:ista step consequence}), holds for any $0 < \eta \leq 1/L$.
    
    Let $\bm{x} = \bm{v} = \bm{x}_k$ then $\bm{p}_{\eta}(\bm{v}) = \bm{x}_{k+1}$ by (\ref{eqn:ISTA update step}). Hence, for fixed step size $\eta_{k+1} = \eta$,
    (\ref{eqn:ista step consequence}) produces (\ref{eqn:ista bound}) with $c = \frac{1}{2\eta}$. Next, when using a backtracking linesearch, observe for $0 < \eta_{k+1} \leq 1/L$ that (\ref{eqn:ista step req}) holds and implies the backtracking linesearch inequality (\ref{eqn:backtracking linesearch ISTA}). Combining (\ref{eqn:ista step consequence}) with the fact that $\eta_{k+1}\leq 1$ produces (\ref{eqn:ista bound}) with $c = 1/2$.
\end{proof}

Lastly, we show that fixed points of ISTA are stationary.
\begin{lemma}\label{lemma:ISTA fixed points are stationary}
    Every ISTA mapping fixed point is stationary.
\end{lemma}
\begin{proof}
    Let $\bm{x}^*$ be a fixed point, i.e. $\bm{x}^* = \textnormal{prox}_{\eta h}(\bm{x}^* - \eta \nabla f(\bm{x}^*))$ for $\eta > 0$. By definition of the proximal operator $\bm{x}^*$ satisfies $2)$ of Definition~\ref{def:optimality conditions} and is a stationary point.
\end{proof}

\subsection{Proposition~\ref{prop:convergence of const function values} Details} \label{apndx:ISTA and GD cost function convergence}

To prove Proposition~\ref{prop:convergence of const function values} we use Lemma~\ref{lemma:PGD step bound}, Lemma~\ref{lemma:ista step bound}, and the following Lemma about sublevel sets.

\begin{lemma}\label{lemma:sublevel set is compact}
    Let $(\mathcal{A}_1)$ hold and $(\mathcal{A}_2)$ or $(\mathcal{A}_3)$ hold.
    The set
    \begin{align*}
   S(\bm{u}_0,\bm{z}_0) = \{(\bm{u},\bm{z})\in \mathbb{R}^n\times \mathfrak{Z}: F(\bm{u},\bm{z})\leq F(\bm{u}_0,\bm{z}_0)\},
\end{align*}
for $(\bm{u}_0,\bm{z}_0)\in\mathbb{R}^n\times\mathfrak{Z}$, is compact.
\end{lemma}
\begin{proof}
    Both ($\mathcal{A}_2$) and ($\mathcal{A}_3$) imply $F$ is continuous  and so $S(\bm{u}_0,\bm{z}_0)$ is closed. By ($\mathcal{A}_1$), for any $i = 1, 2, \ldots, n$,
    \begin{align*}
        \underset{u_i\to\pm\infty}{\lim} F(\bm{u},\bm{z})\to\infty \hspace{.25cm} \textnormal{ and } \hspace{.25cm} \underset{z_i\to\infty}{\lim} F(\bm{u},\bm{z})\to\infty.
    \end{align*}
    Hence $S(\bm{u}_0,\bm{z}_0)$ is bounded and thus is compact.
\end{proof}

We now show $\{F(\bm{u}_k,\bm{z}_k)\}_{k = 1}^\infty$ converges.
\begin{proof}[Proof of Proposition~\ref{prop:convergence of const function values}]
     Let $f(\bm{u},\bm{z}) = \frac{1}{2}\norm{\bm{y}-A(\bm{z}\odot\bm{u})}_2^2$, which is a $C^2$ function in $\bm{z}$. When ($\mathcal{A}_3$) holds, $F(\bm{u},\bm{z})$ is also a $C^2$ function in $\bm{z}$. Thus, $H_{f;\bm{z}}(\bm{u},\bm{z})$ and $H_{F;\bm{z}}(\bm{u},\bm{z})$, the $\bm{z}$ Hessians of $f$ and $F$ respectively, are continuous. Hence, $\norm{H_{f;\bm{z}}(\bm{u},\bm{z})}_2$ and $\norm{H_{F;\bm{z}}(\bm{u},\bm{z})}_2$ are also continuous.  
     
     Let ch$(\mathcal{S})$ denote the convex hull of a set $\mathcal{S}$. By Lemma~\ref{lemma:sublevel set is compact}, $S(\bm{u}_0,\bm{z}_0)$ is compact and thus Carath\'{e}odary's Theorem implies ch$(S(\bm{u}_0,\bm{z}_0))$ is also compact. By the extreme value theorem $\norm{H_{f;\bm{z}}(\bm{u},\bm{z})}_2$ and $\norm{H_{F;\bm{z}}(\bm{u},\bm{z})}_2$ obtain a maximum on ch$(S(\bm{u}_0,\bm{z}_0))$. Therefore, by the mean value theorem $f$ and $F$ have Lipschitz continuous gradient on ch$(S(\bm{u}_0,\bm{z}_0))$ and thus on $S(\bm{u}_0,\bm{z}_0)$ as $S(\bm{u}_0,\bm{z}_0)\subseteq$ ch$(S(\bm{u}_0,\bm{z}_0))$. 
     
     Hence, for PGD G-CG-LS when $(\mathcal{A}_3)$ holds, Lemma~\ref{lemma:PGD step bound} holds. Similarly, for ISTA G-CG-LS where $(\mathcal{A}_2)$ holds, Lemma~\ref{lemma:ista step bound} (taking $r = \mathcal{R}$) holds. Consequently, for all $k\in\mathbb{N}$, any $j = 1, \ldots, J$, and either PGD or ISTA G-CG-LS, we have
    \begin{align}
       \scalebox{.97}{${\displaystyle F(\bm{u}_{k-1},\bm{z}_k^{(j-1)}) - F(\bm{u}_{k-1},\bm{z}_k^{(j)}) \geq c \norm{\bm{z}_k^{(j)}-\bm{z}_k^{(j-1)}}_2^2.}$} \label{eqn:z step bound}
    \end{align}
    Then $F(\bm{u}_{k-1},\bm{z}_{k-1}) \geq F(\bm{u}_{k-1},\bm{z}_k)\geq F(\bm{u}_{k},\bm{z}_k)$ holds for all $k\in\mathbb{N}$. Therefore, $\{F(\bm{u}_k,\bm{z}_k)\}_{k = 1}^\infty$ is a monotonic decreasing sequence that is bounded below and thus converges.
\end{proof}

\subsection{Theorem~\ref{thm:ISTA and PGD convergence} Details}\label{apndx:ISTA and GD estimate convergence}

Proving Theorem~\ref{thm:ISTA and PGD convergence} uses Proposition~\ref{prop:convergence of const function values}, Lemma~\ref{lemma:PGD fixed points are stationary} and~\ref{lemma:ISTA fixed points are stationary}.
\begin{proof}[Proof of Theorem~\ref{thm:ISTA and PGD convergence}]
    Summing (\ref{eqn:z step bound}), from the Proposition~\ref{prop:convergence of const function values} proof, over $j$, and using that $\bm{z}_k^{(0)} = \bm{z}_{k-1}$, $\bm{z}_k^{(J)} = \bm{z}_k$, and $F(\bm{u}_{k-1},\bm{z}_k) \geq F(\bm{u}_{k},\bm{z}_k)$ produces
\begin{align}
\scalebox{0.99}{${\displaystyle F(\bm{u}_{k-1},\bm{z}_{k-1}) - F(\bm{u}_{k}, \bm{z}_k) \geq c\sum_{j = 1}^J \norm{\bm{z}_k^{(j)}-\bm{z}_k^{(j-1)}}_2^2.}$} \label{eqn:z update bound}
\end{align}
By Proposition~\ref{prop:convergence of const function values}, $\{F(\bm{u}_\ell, \bm{z}_{\ell})\}_{\ell = 1}^\infty$ converges and we let $F^*$ be the limit point. Hence, summing (\ref{eqn:z update bound}) over $k \in\mathbb{N}$ and using that the left hand side of (\ref{eqn:z update bound}) is a telescoping sum we have
\begin{align}
    F(\bm{u}_0,\bm{z}_0)-F^* \geq c\sum_{k = 1}^\infty\sum_{j = 1}^J \norm{\bm{z}_k^{(j)}-\bm{z}_k^{(j-1)}}_2^2. \label{eqn:projected GD total bound}
\end{align}
Thus $\underset{k\to\infty}{\lim} \norm{\bm{z}_k^{(j)}-\bm{z}_k^{(j-1)}}_2^2 \to 0$ for every $j = 1, 2, \ldots, J$. 

We make two remarks. First, note that for any real-valued sequence $\{x_k\}_{k = 1}^\infty$ the Cauchy-Schwarz inequality implies $\sum_{j = 1}^J x_j^2 \geq \frac{1}{J}\left(\sum_{j = 1}^J x_j\right)^2.$ Second, using a telescoping sum and the triangle inequality note that
\begin{align*}
   \resizebox{\columnwidth}{!}{${\displaystyle \norm{\bm{z}_k - \bm{z}_{k-1}}_2 = \norm{\sum_{j = 1}^J (\bm{z}_k^{(j)}-\bm{z}_k^{(j-1)})}_2 \leq \sum_{j = 1}^J \norm{\bm{z}_k^{(j)}-\bm{z}_k^{(j-1)}}_2. }$}
\end{align*}
Combining these two notes and (\ref{eqn:projected GD total bound}) produces
\begin{align*}
    F(\bm{u}_0,\bm{z}_0)-F^* \geq \frac{c}{J}\sum_{k = 1}^\infty \norm{\bm{z}_k-\bm{z}_{k-1}}_2^2.
\end{align*}
Hence, it also holds that $\underset{k\to\infty}{\lim} \norm{\bm{z}_k-\bm{z}_{k-1}}_2^2 \to 0$. By continuity of the Tikhonov solution it similarly holds that $\underset{k\to\infty}{\lim} \norm{\bm{u}_k-\bm{u}_{k-1}}_2^2 = \underset{k\to\infty}{\lim} \norm{\mathcal{T}(\bm{z}_k)-\mathcal{T}(\bm{z}_{k-1})}_2^2 \to 0$.

Next, let $(\bm{u}^*,\bm{z}^*)$ be any limit point of $\{(\bm{u}_k,\bm{z}_k)\}_{k = 1}^\infty$ and $\{(\bm{u}_{k_i},\bm{z}_{k_i})\}_{i = 1}^\infty$ the subsequence converging to $(\bm{u}^*,\bm{z}^*)$. By Lemma~\ref{lemma:PGD step bound} and Lemma~\ref{lemma:ista step bound} the sequence of step sizes $\{\eta_{k_i}^{(1)}\}_{i = 1}^\infty$ is bounded and thus there exists a convergent subsequence $\{\eta_{k_{i_\ell}}^{(1)}\}_{\ell = 1}^\infty$. Let $\eta^*$ be the limit point. As $\{(\bm{u}_{k_i},\bm{z}_{k_i})\}_{i = 1}^\infty$ converges every subsequence converges to the same limit point implying $\{(\bm{u}_{k_{i_\ell}},\bm{z}_{k_{i_\ell}})\}_{\ell = 1}^\infty$ converges to $(\bm{u}^*,\bm{z}^*)$. As $\underset{k\to\infty}{\lim} \norm{\bm{z}_k-\bm{z}_{k-1}}_2^2 \to 0$ and $\underset{k\to\infty}{\lim} \norm{\bm{u}_k-\bm{u}_{k-1}}_2^2 \to 0$  then $\{(\bm{u}_{k_{i_\ell}-1},\bm{z}_{k_{i_\ell}-1})\}_{\ell=1}^\infty$ converges to $(\bm{u}^*,\bm{z}^*).$ Similarly, as $\underset{k\to\infty}{\lim} \norm{\bm{z}_k^{(j)}-\bm{z}_k^{(j-1)}}_2^2 \to 0$ for every $j = 1, 2, \ldots, J$ we have  
\begin{align*}
&\underset{\ell\to\infty}{\lim} \norm{\bm{z}_{k_{i_\ell}}^{(1)}-\bm{z}_{k_{i_\ell}}^{(0)}}_2^2 =\underset{\ell\to\infty}{\lim} \norm{\bm{z}_{k_{i_\ell}}^{(1)}-\bm{z}_{k_{i_\ell}-1}}_2^2\to 0, 
\end{align*}
which implies $\{\bm{z}_{k_{i_\ell}}^{(0)}\}_{\ell = 1}^\infty$ and  $\{\bm{z}_{k_{i_\ell}}^{(1)}\}_{\ell = 1}^\infty$ converge to $\bm{z}^*$. In PGD G-CG-LS, by continuity of $\mathcal{P}_{\mathfrak{Z}}$~\cite{cheney1959proximity} and $\nabla_{\bm{z}}F$, observe
\begin{align*}
    \lim_{\ell\to \infty} \bm{z}_{k_{i_\ell}}^{(1)} &= \lim_{\ell\to \infty} \mathcal{P}_{\mathfrak{Z}}\left(\bm{z}_{k_{i_\ell}}^{(0)} - \eta_{k_{i_\ell}}^{(1)}\nabla_{\bm{z}} F(\bm{u}_{k_{i_\ell}-1},\bm{z}_{k_{i_\ell}}^{(0)})\right) \\
    \bm{z}^* &= \mathcal{P}_{\mathfrak{Z}}(\bm{z}^*-\eta^* \nabla_{\bm{z}} F(\bm{u}^*,\bm{z}^*)).
\end{align*}
Therefore, by Lemma~\ref{lemma:PGD fixed points are stationary} it holds that
\begin{align}
    \langle \nabla_{\bm{z}} F(\bm{u}^*, \bm{z}^*), \bm{z}-\bm{z}^*\rangle \geq 0 \hspace{.5cm} \textnormal{ for all } \bm{z}\in \mathfrak{Z}. \label{eqn:pgd z optimiality condition}
\end{align}
In ISTA G-CG-LS, by continuity of the proximal operator \cite{rockafellar1976monotone}
\begin{align*}
  & \resizebox{\columnwidth}{!}{${\displaystyle \lim_{\ell\to\infty} \bm{z}_{k_{i_\ell}}^{(1)} = \lim_{\ell\to\infty} \textnormal{prox}_{\eta_{k_{i_\ell}}^{(1)}\mathcal{R}}\left(\bm{z}_{k_{i_\ell}}^{(0)}-\eta_{k_{i_\ell}}^{(1)} A_{\bm{u}_{k_{i_\ell}-1}}^T(A_{\bm{u}_{k_{i_\ell}-1}}\bm{z}_{k_{i_\ell}}^{(0)} - \bm{y})\right)}$} \\
    & \bm{z}^* = \textnormal{prox}_{\eta^*\mathcal{R}}\left(\bm{z}^* - \eta^* A_{\bm{u}^*}^T(A_{\bm{u}^*}\bm{z}^*-\bm{y})\right).
\end{align*}
Therefore, by Lemma~\ref{lemma:ISTA fixed points are stationary}, since $\partial_{\bm{z}} F(\bm{u},\bm{z}) = \nabla_{\bm{z}} f(\bm{u},\bm{z}) + \partial \mathcal{R}(\bm{z})$, there exists a $\bm{d}^*\in \partial \mathcal{R}(\bm{z}^*)$ such that
\begin{align}
    \left\langle \nabla_{\bm{z}} f(\bm{u}^*,\bm{z}^*)+\bm{d}^*, \bm{z}-\bm{z}^*\right\rangle \geq 0 \hspace{.1cm} \textnormal{for all } \bm{z}\in\mathfrak{Z}. \label{eqn:ista z optimality condition}
\end{align}

As $\bm{u}^*$ is the global minimizer of $F(\bm{u},\bm{z}^*)$ then
\begin{align}
    \langle \nabla_{\bm{u}} F(\bm{u}^*,\bm{z}^*), \bm{u}-\bm{u}^*\rangle &\geq 0 \hspace{.1cm} \textnormal{for all } \bm{u}\in\mathbb{R}^n. \label{eqn:optimiality in u}
\end{align}
Adding together (\ref{eqn:pgd z optimiality condition}) and (\ref{eqn:optimiality in u}), for all $(\bm{u},\bm{z})\in \mathbb{R}^n\times \mathfrak{Z}$
\begin{align*}
    \langle \nabla F(\bm{u}^*,\bm{z}^*),\begin{bmatrix}
        \bm{u} &
        \bm{z}
    \end{bmatrix}^T -\begin{bmatrix}
        \bm{u}^* &
        \bm{z}^*
    \end{bmatrix}^T \rangle &\geq 0
\end{align*}
and, similarly, adding together (\ref{eqn:ista z optimality condition}) and (\ref{eqn:optimiality in u})
\begin{align*}
   \resizebox{\columnwidth}{!}{${\displaystyle \langle \nabla f(\bm{u}^*,\bm{z}^*) + \begin{bmatrix}
        P_{\bm{u}}^{-1}\bm{u}^* &
        \bm{d}^*
    \end{bmatrix}^T ,\begin{bmatrix}
        \bm{u} &
        \bm{z}
    \end{bmatrix}^T -\begin{bmatrix}
        \bm{u}^* &
        \bm{z}^*
    \end{bmatrix}^T\rangle \geq 0. }$}
\end{align*}
Therefore $(\bm{u}^*,\bm{z}^*)$ is a stationary point of $F(\bm{u},\bm{z})$.
\end{proof}

\subsection{Proposition~\ref{prop:convergence to set of const value} and Corollary~\ref{corollary:convergence} Details} \label{apndx:covergence to set of const value}
 
Proving Proposition~\ref{prop:convergence to set of const value} uses Proposition~\ref{prop:convergence of const function values} and Theorem~\ref{thm:ISTA and PGD convergence}.
\begin{proof}[Proof of Proposition~\ref{prop:convergence to set of const value}]
Let $\mathcal{S}$ be the set of stationary points of (\ref{eqn:cost function}) and $\{(\bm{u}_k,\bm{z}_k)\}_{k = 0}^\infty$ the G-CG-LS sequence with limit point set $\mathcal{L}$. By Theorem~\ref{thm:ISTA and PGD convergence}, $\mathcal{L}\subseteq \mathcal{S}$ and $\{(\bm{u}_k,\bm{z}_k)\}_{k = 0}^\infty$ satisfies $\lim_{k\to\infty} \norm{(\bm{u}_{k+1},\bm{z}_{k+1}) - (\bm{u}_k,\bm{z}_k)}_2\to 0$. Hence, from~\cite{lange2013optimization} (Proposition 12.4.1) $\mathcal{L}$ is connected. By Proposition~\ref{prop:convergence of const function values}, $\{F(\bm{u}_k,\bm{z}_k)\}_{k = 1}^\infty$ converges to a value $F^*$. Thus, for every subsequence $\{(\bm{u}_{k_i},\bm{z}_{k_i})\}_{i = 1}^\infty$, $\{F(\bm{u}_{k_i},\bm{z}_{k_i})\}_{i = 1}^\infty$ converges to $F^*$. Hence (\ref{eqn:cost function}) is constant on $\mathcal{L}$ with value $F^*$. Finally, by definition $\{(\bm{u}_k,\bm{z}_k)\}_{k = 0}^\infty$ converges to $\mathcal{L}$, which is closed~\cite{rudin1953principles}.
\end{proof}

Lastly, we prove Corollary~\ref{corollary:convergence} using Proposition~\ref{prop:convergence to set of const value}.
\begin{proof}[Proof of Corollary~\ref{corollary:convergence}]
    By Proposition~\ref{prop:convergence to set of const value} the G-CG-LS sequence $\{(\bm{u}_k,\bm{z}_k)\}_{k = 0}^\infty$ converges to its set of limit points $\mathcal{L}$ which is connected. Since each stationary point is isolated then $\mathcal{L}$ is a single point, to which $\{(\bm{u}_k,\bm{z}_k)\}_{k = 0}^\infty$ converges.
\end{proof}

\section{Acknowledgements}
We would like to thank all anonymous reviewers for their thoughtful comments that helped us improve this manuscript.

\bibliographystyle{IEEEtran}
\bibliography{main}

\end{document}

\setlength\tabcolsep{2.4pt}
\setlength{\headheight}{15.17393pt}

\title{Supplementary Material: Deep Regularized Compound Gaussian Network for Solving Linear Inverse Problems
}
\author[1, 2]{Carter Lyons
}
\author[1]{Raghu G. Raj
}
\author[2]{Margaret Cheney
}
\affil[1]{U.S. Naval Research Laboratory, Washington, D.C.}
\affil[2]{Colorado State University, Fort Collins, CO}

\maketitle
\thispagestyle{fancy}

\section{Additional DR-CG-Net Numerical Results} \label{sec:additional results}

\begin{figure*}[!t]
\centering
\begin{subfigure}{0.32\textwidth}
    \centering
    \includegraphics[width=\textwidth]{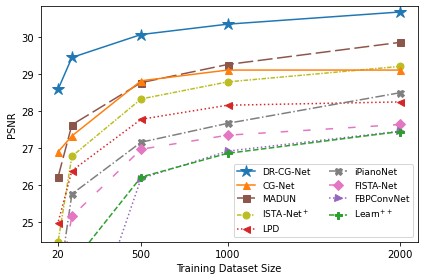}
    \caption{Fifteen uniform angles and 60dB SNR.}
    \label{fig:compare_15_60}
\end{subfigure}
\begin{subfigure}{0.32\textwidth}
    \centering
    \includegraphics[width=\textwidth]{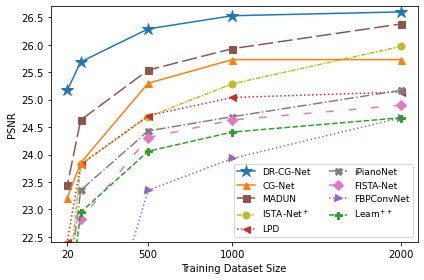}
    \caption{Ten uniform angles and 60dB SNR.}
    \label{fig:compare_10_60}
\end{subfigure}
\begin{subfigure}{0.32\textwidth}
    \centering
    \includegraphics[width=\textwidth]{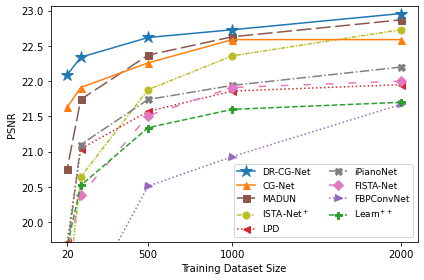}
    \caption{Six uniform angles and 60dB SNR.}
    \label{fig:compare_6_60}
\end{subfigure}

\begin{subfigure}{0.32\textwidth}
    \centering
    \includegraphics[width=\textwidth]{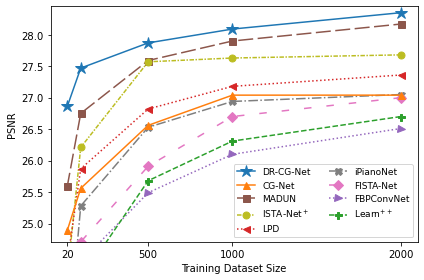}
    \caption{Fifteen uniform angles and 40dB SNR.}
    \label{fig:compare_15_40}
\end{subfigure}
\begin{subfigure}{0.32\textwidth}
    \centering
    \includegraphics[width=\textwidth]{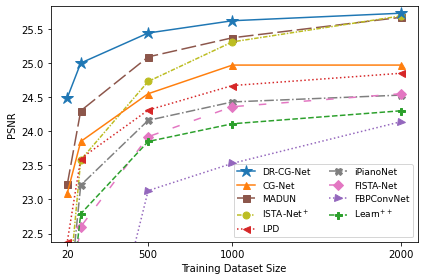}
    \caption{Ten uniform angles and 40dB SNR.}
    \label{fig:compare_10_40}
\end{subfigure}
\begin{subfigure}{0.32\textwidth}
    \centering
    \includegraphics[width=\textwidth]{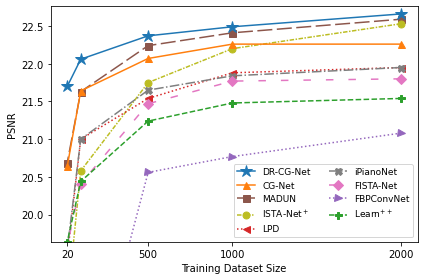}
    \caption{Six uniform angles and 40dB SNR.}
    \label{fig:compare_6_40}
\end{subfigure}
\caption{Average test image reconstruction PSNR when varying the amount of CIFAR10 data in training nine machine learning-based image reconstruction methods. Here, the sensing matrices, $\Psi$, are a Radon transform at 15, 10, or 6 uniformly spaced angles, $\Phi = I$, and the measurement SNR is 60dB or 40dB. \textbf{Our DR-CG-Net method outperforms the compared prior art methods -- for all measurement, noise, and training datasets except the smallest measurement and highest training dataset size -- and does so appreciably in low training}.}
\label{fig:compare}
\end{figure*}

We provide additional simulation details and results for our deep regularized compound Gaussian network (DR-CG-Net). Please refer to the main manuscript for a summary of notation, nomenclature, and a more detailed description of the DR-CG-Net model. Recall that projected gradient descent (PGD) and iterative shrinkage and thresholding (ISTA) scale variable update methods are considered. Thus, PGD DR-CG-Net and ISTA DR-CG-Net, are two variants of DR-CG-Net that correspond to using a PGD or ISTA update of $\bm{z}$, respectively.

We compare DR-CG-Net against ten state-of-the-art methods: (i) compound Gaussian network (CG-Net), (ii) memory augmented deep unfolding network (MADUN), (iii) ISTA-Net$^+$, (iv) FISTA-Net, (v) iPiano-Net, (vi) ReconNet, (vii) LEARN$^{++}$, (viii) Learned Primal-Dual (LPD), (ix) FBPConvNet, and (x) iRadonMAP. Note, LEARN$^{++}$, LPD, FBPConvNet, and iRadonMAP are CT-specific reconstruction methods that rely on the structure of the CT sinogram measurements. On the other hand, MADUN, ISTA-Net$^+$, FISTA-Net, iPiano-Net, and ReconNet are reconstruction methods with particular application in image compressive sensing. Furthermore, we remark that CG-Net, MADUN, ISTA-Net$^+$, FISTA-Net, iPiano-Net, LEARN$^{++}$, and LPD are DNNs formed by algorithm unrolling while ReconNet, iRadonMAP, and FBPConvNet are instead standard DNNs.

For data we use $32\times 32$ CIFAR10 images, $64\times 64$ CalTech101 images, and $128\times 128$ LoDoPaB-CT images. Each image is converted to a single-channel grayscale image, scaled down by the maximum pixel value, and vectorized. We consider a Radon transform and random Gaussian matrix as measurement operators. To each measurement white noise is added producing noisy measurements, $\bm{y}$, at a specified SNR.

\subsection{DR-CG-Net Setup}

Recall that for the unrolled iterations of each DR-CG-Net we set $(K, J) = (3, 4)$ and $(K, J) = (1, 24)$ for $32\times 32$ images and $64\times 64$ or $128\times 128$ images, respectively. Additionally, each $\mathcal{V}_k^{(j)}$ is taken to be a CNN with depth $D = 8$. These network size parameters were chosen empirically such that the time to complete a signal reconstruction was reasonably quick while still producing excellent reconstructions on a validation dataset. Every convolution uses a $3\times 3$ kernel initialized according to the Glorot Uniform distribution with ReLU activation functions and $f_1 = \cdots = f_7 = 32$ and $f_8 = 1$ filter channels.

We initialize each step size as $\delta_k^{(j)} = 1$, set $\gamma_{\max} = 1$, and fix $\epsilon = 10^{-4}.$  Each $\bm{u}$ covariance matrix structure is initialized as a diagonal matrix with $0.1$ or $10$ on the diagonal for Radon transform or Gaussian measurements, respectively. Additionally, in the initial $\bm{z}$ estimate, we set $\mathcal{P}_{[0,b]^n} = \mathcal{P}_{[0, 10]^n}.$ For training datasets of size $N_s = 20$ and $N_s = 100$, we set a scaled identity $\bm{u}$ covariance matrix structure and a tridiagonal $\bm{u}$ covariance matrix structure is used for all other training datasets. Lastly, we train DR-CG-Net using the mean absolute error function with a learning rate of $10^{-4}$ for 2000 epochs.

\subsection{Numerical Results}
For each set of data, we train a PGD DR-CG-Net, ISTA DR-CG-Net, and each of the ten comparison methods. For CIFAR10 images, the training datasets consist of $N_s  = 20, 100, 500, 1000,$ and $2000$ data samples and, after training, 8000 test data samples are provided to every network to assess its performance. For CalTech101 images, the training datasets consist of $N_s = 20$ samples and, after training, 200 test data samples are provided to every network. Average PSNR and SSIM quality metrics on the test dataset reconstructions are used to evaluated network performance where higher values of these metrics correspond to reconstructed images that more closely match the original.

\begin{figure*}[!t]
\centering
    \centering
\begin{subfigure}{0.4\textwidth}
    \centering
    \includegraphics[width=\textwidth]{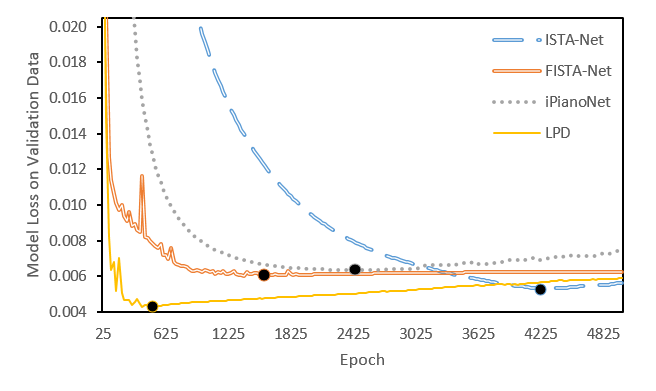}
\end{subfigure}
\begin{subfigure}{0.4\textwidth}
    \centering
    \includegraphics[width=\textwidth]{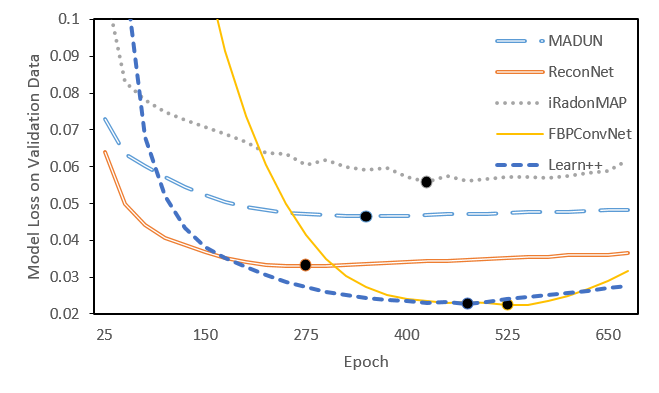}
\end{subfigure}
\begin{subfigure}{0.4\textwidth}
    \centering
    \includegraphics[width=\textwidth]{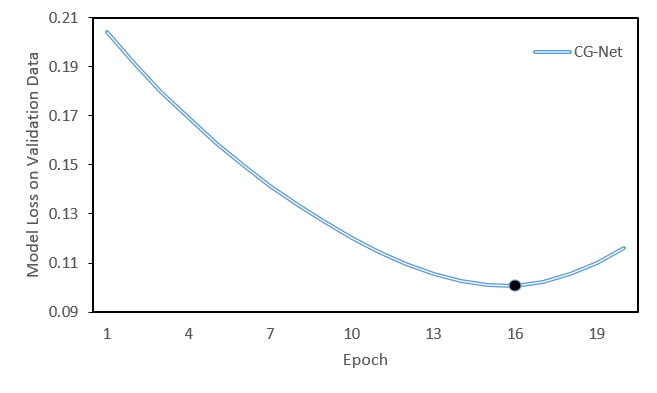}
\end{subfigure}
\begin{subfigure}{0.4\textwidth}
    \centering
    \includegraphics[width=\textwidth]{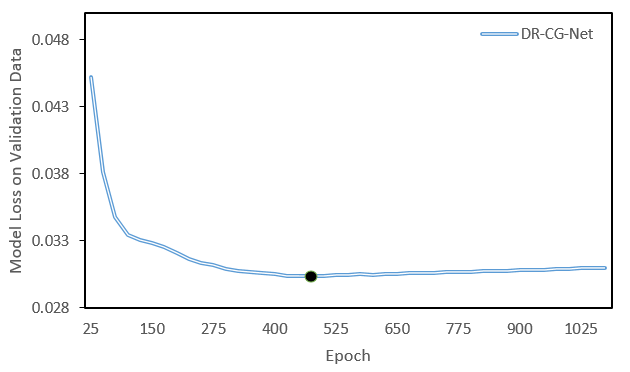}
\end{subfigure}
\caption{\textcolor{black}{Model loss curves on a validation dataset when training each deep learning-based method on Radon inversion from 15 uniformly spaced angles with 20 training samples. The point on each model's loss curve represents the epoch in which that model achieved its best loss on a validation dataset and overfits on subsequent epochs. Only these best loss epoch results are presented for each method. Note, for the plots above, CG-Net uses an SSIM loss function, MADUN and DR-CG-Net use a mean absolute error loss function, and all other methods use a mean square error loss function (possibly with some additional regularization). As the network loss functions are not equivalent, these plots do not contribute a comparison between the methods and only illustrate that each method is maximally trained for every supplied training dataset.}}
\label{fig:loss curves}
\end{figure*}

We remark that every method was trained using early stopping. That is, as shown in Fig. \ref{fig:loss curves}, training was conducted until the model initially overfits as compared to a validation dataset. In doing so, we ensure every model is sufficiently trained while also not being over trained thereby presenting the best performance for each model given the provided set of training data.

Shown in Fig. \ref{fig:compare} is the average PSNR quality over a set of test CIFAR10 image reconstructions from Radon transform measurements for ISTA DR-CG-Net and the comparison methods when each is trained on varying amounts of training data. We note that ReconNet, iRadonMAP, and some instances of FBPConvNet perform significantly lower and are thus omitted from Fig. \ref{fig:compare}. Additionally, as PGD DR-CG-Net performed nearly identically to ISTA DR-CG-Net, it too was omitted from Fig. \ref{fig:compare}. We can see from Fig. \ref{fig:compare} that our DR-CG-Net outperforms all comparative methods in every training scenario, and does so significantly in the lowest training scenarios of 20 and 100 training samples. 

\begin{figure*}
\centering
\begin{subfigure}{.19\linewidth}
    \centering
    \includegraphics[width=\linewidth]{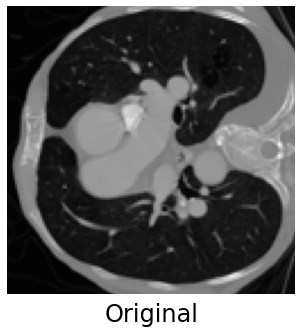}
\end{subfigure}
\begin{subfigure}{.19\linewidth}
    \centering
    \includegraphics[width=\linewidth]{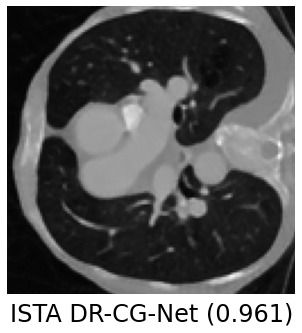}
\end{subfigure}
\begin{subfigure}{.19\linewidth}
    \centering
    \includegraphics[width=\linewidth]{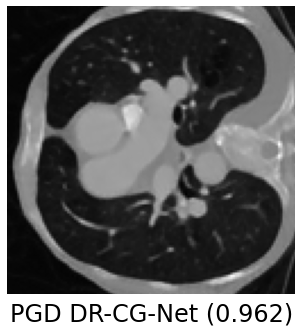}
\end{subfigure}
\begin{subfigure}{.19\linewidth}
    \centering
    \includegraphics[width=\linewidth]{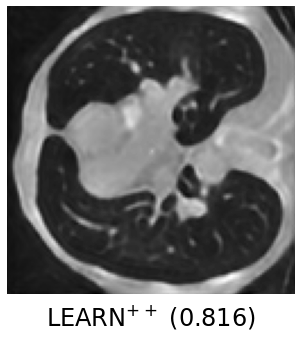}
\end{subfigure}
\begin{subfigure}{.19\linewidth}
    \centering
    \includegraphics[width=\linewidth]{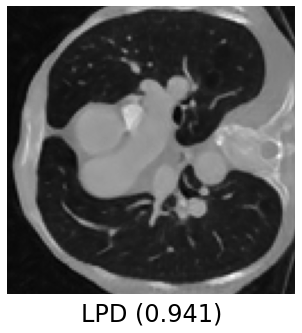}
\end{subfigure}

\begin{subfigure}{.19\textwidth}
    \centering
    \includegraphics[width=\linewidth]{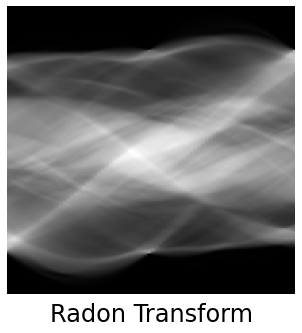}
\end{subfigure}
\begin{subfigure}{.19\linewidth}
    \centering
    \includegraphics[width=\linewidth]{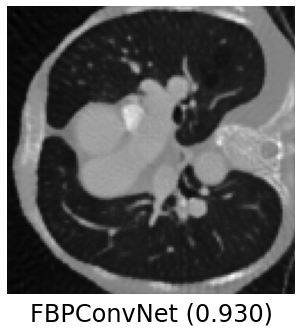}
\end{subfigure}
\begin{subfigure}{.19\linewidth}
    \centering
    \includegraphics[width=\linewidth]{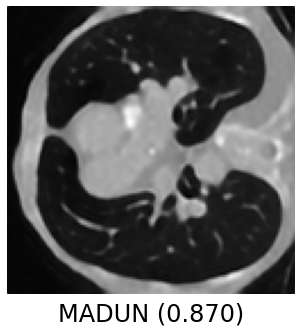}
\end{subfigure}
\begin{subfigure}{.19\linewidth}
    \centering
    \includegraphics[width=\linewidth]{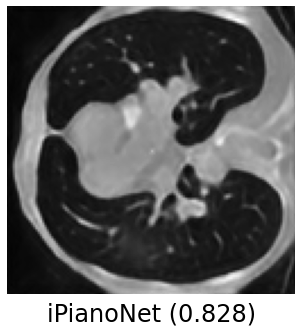}
\end{subfigure}
\begin{subfigure}{.19\linewidth}
    \centering
    \includegraphics[width=\linewidth]{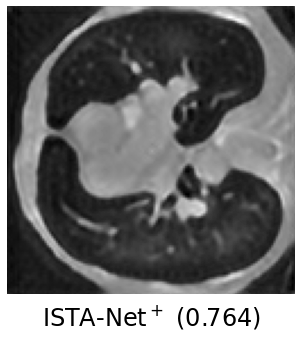}
\end{subfigure}
\caption{Image reconstructions (SSIM) using our DR-CG-Net and six competitive deep learning methods on a $128\times 128$ test scan after training with only 20 samples. The sensing matrix, $\Psi$, is a Radon transform at 76 uniformly spaced angles, $\Phi = I$, and each measurement has an SNR of 60dB. Our DR-CG-Net methods perform best visually and by SSIM.}
\label{fig:radon reconstructions128}
\end{figure*}

\begin{figure*}
\centering
\begin{subfigure}{.19\linewidth}
    \centering
    \includegraphics[width=\linewidth]{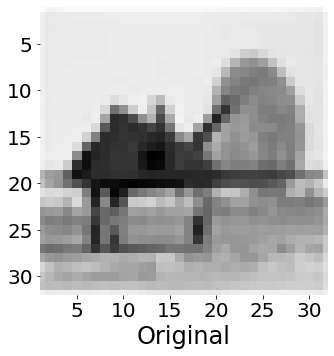}
\end{subfigure}
\begin{subfigure}{.19\linewidth}
    \centering
    \includegraphics[width=\linewidth]{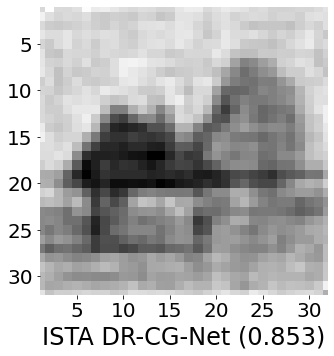}
\end{subfigure}
\begin{subfigure}{.19\linewidth}
    \centering
    \includegraphics[width=\linewidth]{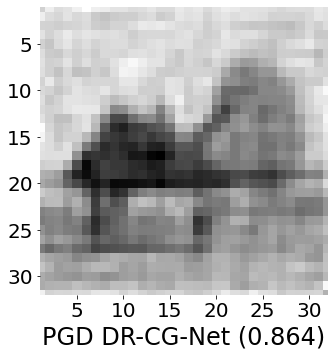}
\end{subfigure}
\begin{subfigure}{.19\linewidth}
    \centering
    \includegraphics[width=\linewidth]{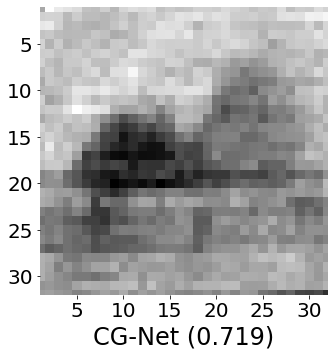}
\end{subfigure}
\begin{subfigure}{.19\linewidth}
    \centering
    \includegraphics[width=\linewidth]{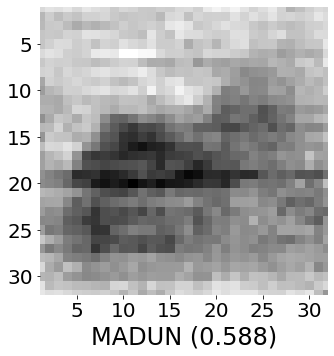}
\end{subfigure}

\begin{subfigure}{.19\textwidth}
    \centering
    \includegraphics[width=\linewidth]{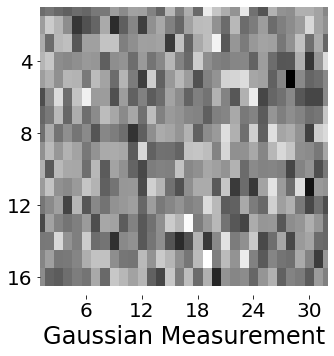}
\end{subfigure}
\begin{subfigure}{.19\linewidth}
    \centering
    \includegraphics[width=\linewidth]{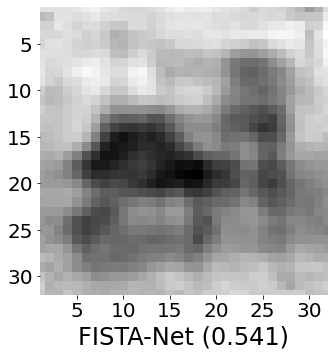}
\end{subfigure}
\begin{subfigure}{.19\linewidth}
    \centering
    \includegraphics[width=\linewidth]{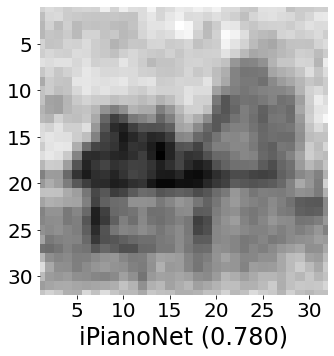}
\end{subfigure}
\begin{subfigure}{.19\linewidth}
    \centering
    \includegraphics[width=\linewidth]{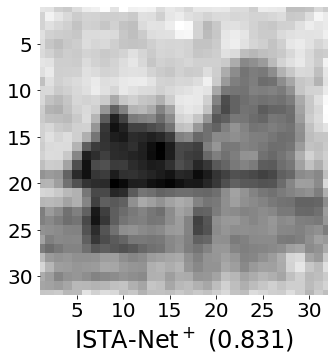}
\end{subfigure}
\begin{subfigure}{.19\linewidth}
    \centering
    \includegraphics[width=\linewidth]{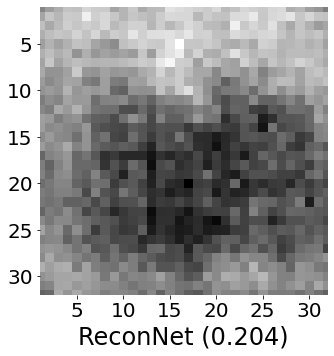}
\end{subfigure}
\caption{Image reconstructions (SSIM) using our DR-CG-Net and six comparative deep learning methods on a $32\times 32$ plane image after training on only 20 samples. The sensing matrix, $\Psi$, is a Gaussian matrix at 0.5 sampling ratio, $\Phi = $ discrete cosine transformation, and each measurement has an SNR of 60dB. Our DR-CG-Net methods perform best visually and by SSIM.}
\label{fig:0.5 ratio reconstructions}
\end{figure*}

Visual comparisons of all methods after training on $20$ LoDoPaB-CT and CIFAR10 samples are provided in Fig.~\ref{fig:radon reconstructions128} and Fig.~\ref{fig:0.5 ratio reconstructions}, respectively. In particular, Fig.~\ref{fig:radon reconstructions128} displays the estimates of a $128\times 128$ scan from 76 uniformly spaced angle Radon transforms with 60dB SNR. Furthermore, Fig.~\ref{fig:0.5 ratio reconstructions} shows the reconstruction of a test plane image from Gaussian transform measurements at 0.5 sampling ratio with 60dB SNR. We observe both visually and quantitatively, by SSIM, that our PGD and ISTA DR-CG-Net methods perform comparably and outperform each prior art method. 

\begin{figure*}[!t]
\centering
\begin{subfigure}{\textwidth}
    \centering
\begin{subfigure}{.32\linewidth}
    \centering
    \includegraphics[width=\linewidth]{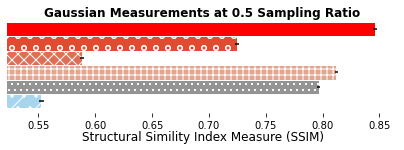}
\end{subfigure}
\begin{subfigure}{.32\linewidth}
    \centering
    \includegraphics[width=\linewidth]{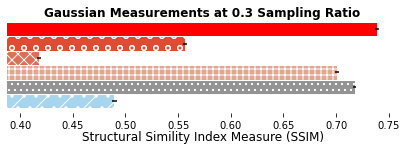}
\end{subfigure}
\begin{subfigure}{.32\linewidth}
    \centering
    \includegraphics[width=\linewidth]{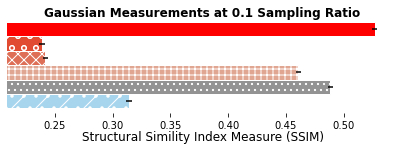}
\end{subfigure}
\begin{subfigure}{.32\linewidth}
    \centering
    \includegraphics[width=\linewidth]{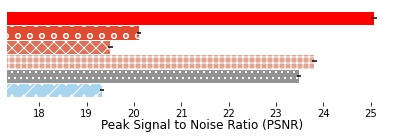}
\end{subfigure}
\begin{subfigure}{.32\linewidth}
    \centering
    \includegraphics[width=\linewidth]{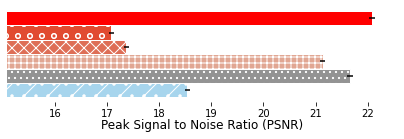}
\end{subfigure}
\begin{subfigure}{.32\linewidth}
    \centering 
    \includegraphics[width=\linewidth]{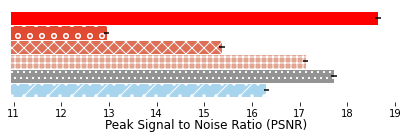}
\end{subfigure}
    \caption{Average SSIM and PSNR, with $99\%$ confidence intervals, for six deep learning-based image estimation methods reconstructing 8000, $32\times 32$ CIFAR10 images from Gaussian measurements with an SNR of 60dB and $\Phi = $ discrete cosine transformation.}
    \label{fig:low training CS}
\end{subfigure}
\begin{subfigure}{.32\linewidth}
    \centering 
    \includegraphics[width=\linewidth]{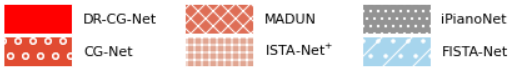}
\end{subfigure}
\caption{Test image reconstruction quality for deep learning-based image estimation methods trained on only 20 samples. \textbf{In all cases, our method, DR-CG-Net given by the top red bar, outperforms the other approaches.}}
\label{fig:low training}
\end{figure*}

Lastly, Fig. \ref{fig:low training} provides average SSIM and PSNR performance on reconstructed CIFAR10 images after training on 20 samples. Each image is reconstructed from random Gaussian measurements at a 0.5, 0.3, or 0.1 sampling ratio with 60dB SNR. We remark that DR-CG-Net significantly outperforms each compared deep learning-based CS method in these low training CS experiments.

%%%%%%%%%%%%%%%%%%%% 128x128 Radon %%%%%%%%%%%%%%%%%
%%%%%%%%%%%%%%%%%%%%%%%%%%%%%%%%%%%%%%%%%%%%%%

%%%%%%%%%%%%%%%%%%%%% 32x32 Gaussian %%%%%%%%%%%%%%%%%
%%%%%%%%%%%%%%%%%%%%%%%%%%%%%%%%%%%%%%%%%%%%%%%

%%%%%%%%%%%%%%%%%%%%%%%%%%%%%%%%%%%%%%%%%%%%%%%%%%%%%%%%%%%%%%%%%%%%

%%%%%%%%% Loss Curve Plots %%%%%%%%%%
%%%%%%%%%%%%%%%%%%%%

\section{Algebraic Proof Manipulations} \label{apndx:proofs}

\subsection{Proposition \ref{main-prop:MAP_estimate} Details}

\begin{proof}[Proof of Proposition \ref{main-prop:MAP_estimate}]
     For random vector $\bm{x}$ let $p_{\bm{x}}(\bm{x})$ be its probability density function and $p_{\bm{x}|\bm{v}}(\bm{x}|\bm{v})$ the conditional probability density given $\bm{v}$. Additionally, let $\mathcal{N}(\bm{\mu},\Sigma)$ be a multivariate normal distribution of mean $\bm{\mu}$ and covariance $\Sigma$. Observe
\begin{align*}
    p_{\bm{y}|\bm{u},\bm{z}}(\bm{y}|\bm{u},\bm{z}) \sim \mathcal{N}(A(\bm{z}\odot\bm{u}), \sigma^2 I).
\end{align*}
Since $\bm{u}\sim\mathcal{N}(\bm{0}, \Sigma_u)$ then using independence of $\bm{u}$ and $\bm{z}$ the MAP estimate of $\bm{z}$ and $\bm{u}$ from equation (\ref{main-eqn:linear_msrmt}) is
\begin{align*}
    &\underset{\bm{u},\bm{z}}{\arg\max}\,\, p_{\bm{u},\bm{z}|\bm{y}}(\bm{u},\bm{z}|\bm{y}) \\
    &= \underset{\bm{u},\bm{z}}{\arg\min}\,\, -\ln(p_{\bm{y}|\bm{u},\bm{z}}(\bm{y}|\bm{u},\bm{z}))-\ln(p_{\bm{u},\bm{z}}(\bm{u},\bm{z})) \\
    &= \underset{\bm{u},\bm{z}}{\arg\min}\,\, -\ln(p_{\bm{y}|\bm{u},\bm{z}}(\bm{y}|\bm{u},\bm{z}))-\ln(p_{\bm{u}}(\bm{u})) -\ln(p_{\bm{z}}(\bm{z})) \\
    &= \underset{\bm{u},\bm{z}}{\arg\min}\,\, \frac{1}{2\sigma^2}||\bm{y} - A(\bm{z}\odot\bm{u})||_2^2 + \frac{1}{2}\bm{u}^T \Sigma_u^{-1}\bm{u} -\ln(p_{\bm{z}}(\bm{z}) \\
    &= \underset{\bm{u},\bm{z}}{\arg\min}\,\, \frac{1}{2}||\bm{y} - A(\bm{z}\odot\bm{u})||_2^2 + \frac{1}{2}\bm{u}^T P_u^{-1}\bm{u} +\mathcal{R}(\bm{z})
\end{align*}
where $\mathcal{R}(\bm{z}) = -\sigma^2\ln(p_{\bm{z}}(\bm{z}))$ and $P_u = \sigma^{-2}\Sigma_u$.
\end{proof}

\subsection{Improving the Tikhonov Solution} \label{apndx:tikhonov solution}
Here, we show the Tikhonov solution in equations (\ref{main-eqn:Tikhonov Solution}) and (\ref{main-eqn:Tikhonov solution rewritten}) are equivalent. Observe using the Woodbury matrix identity
\begin{align*}
    \mathcal{T}(\bm{z}) &= (A_{\bm{z}}^TA_{\bm{z}} + P_u^{-1})^{-1}A_{\bm{z}}^T\bm{y} \\
    &= \left(P_u - P_uA^T_{\bm{z}}(I + A_{\bm{z}}P_uA^T_{\bm{z}})^{-1}A_{\bm{z}}P_u\right)A^T_{\bm{z}}\bm{y} \\
    &= \left(P_uA^T_{\bm{z}} - P_uA^T_{\bm{z}}(I + A_{\bm{z}}P_uA^T_{\bm{z}})^{-1}A_{\bm{z}}P_uA^T_{\bm{z}}\right)\bm{y} \\
    &= P_uA^T_{\bm{z}}\left(I - (I + A_{\bm{z}}P_uA^T_{\bm{z}})^{-1}A_{\bm{z}}P_uA^T_{\bm{z}}\right)\bm{y} \\
    &= \scalebox{.96}{${\displaystyle  P_uA^T_{\bm{z}}(I + A_{\bm{z}}P_uA^T_{\bm{z}})^{-1}\left(I + A_{\bm{z}}P_uA^T_{\bm{z}} - A_{\bm{z}}P_uA^T_{\bm{z}}\right)\bm{y} }$} \\
    &= P_uA^T_{\bm{z}}(I + A_{\bm{z}}P_uA^T_{\bm{z}})^{-1}\bm{y}.
\end{align*}